\documentclass[lettersize,journal]{IEEEtran}
\usepackage{amsmath,amsfonts}
\usepackage{amsthm} 
\usepackage{algorithmic}
\usepackage{algorithm}
\usepackage{array}
\usepackage[caption=false,font=normalsize,labelfont=sf,textfont=sf]{subfig}
\usepackage{textcomp}
\usepackage{stfloats}
\usepackage{url}
\usepackage{verbatim}
\usepackage{graphicx}
\usepackage{cite}
\usepackage{gensymb}
\usepackage[normalem]{ulem}

\usepackage{multirow}
\usepackage{color, colortbl}

\usepackage[english]{babel}

\usepackage{xcolor}

\newtheorem{theorem}{Theorem}
\newtheorem{lemma}[theorem]{Lemma}

\begin{document}

\title{Joint Optimization of {Production and Maintenance} in Offshore Wind Farms: Balancing the Short- and Long-Term Needs of Wind Energy Operation}

\author{Petros Papadopoulos$^1$, Farnaz Fallahi$^2$, Murat Yildirim$^2$, Ahmed Aziz Ezzat$^{3,\dagger}$

\thanks{$^1$ P. Papadopoulos is with ThermoVault in Leuven, Belgium} 
\thanks{$^2$ F. Fallahi and M. Yildirim are with the Department of Industrial \& Systems Engineering at Wayne State University}
\thanks{$^3$ A. Aziz Ezzat is with the Department of Industrial \& Systems Engineering at Rutgers, The State University of NJ}
\thanks{$^\dagger$ Corresponding author: Ahmed Aziz Ezzat, aziz.ezzat@rutgers.edu}
}

\markboth{Paper accepted, August~2023}
{Shell \MakeLowercase{\textit{et al.}}: A Sample Article Using IEEEtran.cls for IEEE Journals}


\maketitle

\begin{abstract}
The rapid increase in scale and sophistication of offshore wind (OSW) farms poses a critical challenge related to the cost-effective operation and management of wind energy assets. A defining characteristic of this challenge is the economic trade-off between two concomitant processes: power production (the primary driver of short-term revenues), and asset degradation (the main determinant of long-term expenses). Traditionally, approaches to optimize production and maintenance in wind farms have been conducted in isolation. In this paper, we conjecture that a joint optimization of those two processes, achieved by rigorously modeling their short- and long-term dependencies, can unlock significant economic benefits for wind farm operators. In specific, we propose a decision-theoretic framework, rooted in stochastic optimization, which seeks a sensible balance of how wind loads are leveraged to harness short-term electricity generation revenues, versus alleviated to hedge against longer-term maintenance expenses. Extensive numerical experiments using real-world data confirm the superior performance of our approach, in terms of several operational performance metrics, relative to methods that tackle the two problems in isolation. 
\end{abstract}


\section{Introduction} \label{S:intro}
\IEEEPARstart{D}{espite} its rapid growth, the large-scale penetration of offshore wind (OSW) into modern-day power systems is contingent on innovative solutions to reduce its operations and maintenance (O\&M) expenditures. To wind farm operators, lowering O\&M costs hinges on solutions to two major challenges: (1) Maintenance scheduling\textemdash to minimize the long-term maintenance requirements of OSW energy assets, and (2) Production optimization\textemdash to maximize the short-term electricity generation revenues for a fleet of OSW turbines.

Solutions to the first challenge primarily seek optimal maintenance schedules by monitoring, and further {predicting}, the degradation status of wind energy assets, and then determining when and where to commit a repair action. The ultimate goal is to prolong the assets' useful life and reduce their long-term maintenance expenses \cite{besnard2011stochastic,yildirim2017integrated,fallahi2022chance, papadopoulos2022seizing}. {There has been a rich literature on maintenance optimization for (offshore) wind farms, which can be broadly grouped into two main clusters: time-based (TBS) and condition-based (CBS) strategies. TBS refer to periodic policies that identify an optimal window (or frequency) for maintenance decisions \cite{ding2012opportunistic}. CBS, on the other hand, heavily rely on real-time sensory data to inform maintenance scheduling decisions \cite{byon2010optimal, byon2010season}, and are often linked to the concept of ``smart maintenance'' in the broader context of asset management \cite{alvarez2022power}.}

In parallel to maintenance optimization, the second challenge, i.e. production optimization, 
mostly entails turbine control strategies that maximize the short-term power capture of a fleet of wind turbines (WTs) \cite{yang2021review}. {Based on the control mechanism, those include pitch \cite{gebraad2013model}, torque \cite{johnson2012assessment}, and yaw control \cite{fleming2014evaluating}.} {Yaw-based optimization, which is the focus of this work, adjusts the orientation of the WT's rotor to maximize its power production, and has been shown to increase turbine- and farm-level production by up to $13$\%}\cite{kragh2015potential, song2018maximum, howland2019wind}. 

To date, those two operational challenges, namely maintenance and production optimization, have been mostly addressed in isolation, perhaps due to the different time scales at which their respective decisions are made. In reality, however, those two problems are intrinsically coupled: Short-term production decisions have long-term maintenance implications, and vice-versa. On one hand, the overwhelming majority of turbine control strategies often prioritize short-term power gains with little (or no) consideration of the accumulating structural degradation (and ultimately, maintenance expenses) which unfold due to those turbine control actions. On the other hand, maintenance optimization strategies, by and large, optimize for long-term, production-agnostic maintenance objectives, potentially forfeiting substantial, shorter-term production gain opportunities. This trade-off gives rise to an under-explored research question addressed in this work: \textit{Can we propose a decision-theoretic framework that balances how wind loads are leveraged to harness short-term electricity generation revenues, versus alleviated to hedge against longer-term failure risks and maintenance expenses?} 

In response, we propose POSYDON (short for \textit{\underline{p}roduction-\underline{o}ptimized \underline{s}tochastic opportunistic maintenance scheduler with \underline{y}aw \underline{d}ecision c\underline{on}trol}). Primarily rooted in stochastic programming, POSYDON jointly optimizes production and maintenance decisions in tandem, thereby balancing the short- and long-term financial targets of wind farm operation. In doing so, POSYDON equips the operators with the capability to either prioritize short-term production gains relative to long-term maintenance savings when economically justified, or conversely, de-prioritize short-term production gains to hedge against the failure risks 
of critically degraded 
assets. 

\IEEEpubidadjcol 

{There have been a number of recent research efforts to couple maintenance and production decisions in (offshore) wind farms. In the maintenance optimization literature, the so-called ``opportunistic maintenance'' is a class of maintenance models which optimally groups maintenance actions, primarily to share setup costs, but also to optimize revenue- and availability-related objectives \cite{papadopoulos2022seizing, papadopoulos2022stochos}. For example, \cite{yang2020operations, yang2020operations2} propose opportunistic models that consider the impact of maintenance opportunities on power production and revenues. Along the same line, \cite{erguido2017dynamic} propose a dynamic scheduling approach with variable reliability thresholds to balance short-term production revenues and long-term reliability objectives. Coupling maintenance and economic dispatch decisions has also been proposed to maximize short- and long-term profitability of OSW farms \cite{mazidi2017strategic}. Yet, none of those efforts explicitly model short-term production control decisions (e.g., yaw optimization), let alone the implications of such decisions on longer-term asset health, and ultimately, on the pursuant maintenance schedules, as we propose herein.} 

{In parallel, there is an active line of research in the production optimization literature on yaw-based load alleviation, wherein non-zero yaw misalignment (YM) is introduced to balance the power production and structural loading acting on critical wind turbine components \cite{kragh2014load, damiani2018assessment, van2016yaw}. Our work extends those emerging efforts by going beyond short-term load alleviation objectives and instead, rigorously modeling
the long-term maintenance implications (i.e., failure risks and costs) incurred by those short-term control strategies.} 

In summary, the contributions of this work are as follows: 
\begin{itemize}
\item[($\mathcal{C}1$)] We propose a stochastic mixed integer linear programming (MILP) formulation, referred to as POSYDON, which takes probabilistic forecasts about environmental and operational parameters (e.g., wind, wave, electricity prices, asset degradation), and outputs a set of (long-term) maintenance schedules and (short-term) yaw control decisions, which jointly maximize the OSW farm profitability. 
\item[($\mathcal{C}2$)] To achieve $\mathcal{C}1$, we propose a novel predictive model to forecast the long-term degradation and failure risk of a WT, which uniquely links the structural loads incurred by short-term yaw decisions to their long-term maintenance implications. Similarly, we rely on a state-of-the-art predictive model to estimate a WT's power output as a function of yaw decisions (among other variables). Combined, those two predictive models offer a unique predictive-prescriptive coupling\textemdash where decisions impact predictions and vice-versa\textemdash thereby enabling the degradation and wind power uncertainties to be modeled as \textit{decision-dependent} entities in POSYDON \cite{hellemo2018decision}. 
\item[($\mathcal{C}3$)] {We derive key insights, analyses, and findings that demonstrate how POSYDON, through its rigorous modeling of short- and long-term decisions, outperforms single-faceted strategies that separately optimize production or maintenance, across several O\&M metrics, including downtime, utilization, and total costs. Those insights are derived through extensive experiments on real-world data
from the United States (US) NY/NJ Bight\textemdash where several GW-scale projects are in-development. We believe that those insights and analyses can be directly useful to the rising OSW sector in the US and elsewhere.} 
\end{itemize}

The remainder of this paper is structured as follows. In Section \ref{S:YM_degr}, we present our yaw-dependent degradation model to predict the failure risks in WT assets. In Section \ref{S:Opt_model}, we present the mathematical formulation of POSYDON. In Section \ref{S:numerical_exp}, we describe our experimental setup, along with the optimization results.
We then conclude the paper in Section \ref{S:concl}.

\section{Yaw-dependent Degradation Modeling} \label{S:YM_degr}
There is an active literature on predictive modeling to forecast the health condition of WTs \cite{rezamand2020critical}. The essence of those so-called ``\textit{degradation models}'' is to regard the data collected by turbine-mounted sensors (e.g., vibration, temperature) as \textit{degradation signals} which are used to predict the remaining useful life (RUL) of a WT. Our contribution herein is to propose the first degradation model that rigorously considers the impact of short-term, yaw-induced loading on the RUL predictions of a WT. Those ``yaw-dependent'' RUL predictions will be used as input to the optimization model in Section III, thereby enabling the linkage between the short- and long-term production and maintenance decisions addressed in this work.   

In specific, we propose a two-stage degradation modelling approach. In the first stage, we formulate a \textit{baseline degradation model} for a WT operating under ``baseline'' (or nominal) conditions. Those baseline conditions represent the typical loading incurred by a WT during its normal operation. This is discussed in Section II.A. Understandably, WTs do not often experience the baseline conditions, but instead are subjected to time-varying loads, due to, on one hand, the change in the WT's environment (e.g., variations in wind speeds), and on the other hand, due to the yaw control decisions, which either increase or decrease the loads relative to baseline conditions. Hence, in the second stage, the baseline degradation model is expanded to consider the impact of those time-varying loads through a \textit{time-transformed degradation model}, which maps the baseline degradation level (under constant loading) into an ``actual'' degradation level (considering time-varying loading). This is presented in detail in Section II.B. 

Without loss of generality, we focus on 
degradation in WT blades. Blades are major contributors to the downtime and maintenance requirements in OSW farms, with repair costs that can reach up to \$$95$K/failure \cite{carroll2016failure}. We specifically model the blade root flapwise bending moment (FBM), which is a main driver of fatigue-induced blade failures \cite{ennis2018wind}. In addition to degradation signals, we consider the wind speed and yaw misalignment (YM) as inputs to our predictive degradation model, as both of those variables are known to highly impact FBM \cite{bergami2014analysis,kragh2014load, hure2015optimal, kragh2015potential}. Hereinafter, we denote wind speed and yaw misalignment as $\nu$ and $\tilde{\boldsymbol{\gamma}}$, respectively. 

\subsection{Baseline degradation model} \label{S:nom_deg_model}
The baseline model describes the blade degradation under constant nominal loading. We define nominal conditions as (\textit{i}) mean wind speed $\nu^0$, and (\textit{ii}) perfect yaw misalignment $\tilde{\gamma}^0$. We call the random process by which the blade degrades under baseline conditions as the ``baseline degradation process,'' which is denoted by $\{A^0(t);t\geq0\}$, and modeled as in (\ref{eq:degradation_function}).
\begin{equation} \label{eq:degradation_function}
{A}^0(t)=\alpha^0+\beta^0t+\epsilon(t;\sigma),
\end{equation}
where $\alpha^0$ and $\beta^0$ represent the initial amplitude and the baseline degradation rate, respectively. The term $\epsilon(t;\sigma)$ is Brownian error, i.e. $\epsilon(t) = \sigma W(t)$, denoting the process noise. {The functional form in \eqref{eq:degradation_function} is 
is a prevalent choice in the parametric degradation modeling literature, and has been shown to effectively describe the degradation of a broad range of failure processes \cite{zhang2018degradation,gebraeel2006sensory,zhou2016novel, zhai2017rul}.} 

Given a set of observed degradation signals collected during the WT operation,  $\boldsymbol{d}_{t}=(d_1,...,d_{t})$, we can then follow a Bayesian approach to continuously update the degradation model parameters, in light of $\boldsymbol{d}_t$. Letting $\pi(\alpha^0)$ and $\pi(\beta^0)$ denote the prior distributions of $\alpha^0$ and $\beta^0$ respectively, then the joint posterior distribution of $\alpha^0$ and $\beta^0$, denoted by $\pi'(\alpha^0, \beta^0)$ can be estimated as in (\ref{eq:degradation_function2}). 
\begin{equation} \label{eq:degradation_function2}
\pi^{'}(\alpha^0,\beta^0)= P(\boldsymbol{d}_{t}|\alpha^0,\beta^0) \pi(\alpha^0)\pi(\beta^0)/P(\boldsymbol{d}_{t}),
\end{equation}
where $P(\boldsymbol{d}_{t}|\alpha^0,\beta^0)$ is the likelihood function, and $P(\boldsymbol{d}_{t})$ is a normalization constant. If we assume Gaussian priors for $\pi(\alpha^0)$ and $\pi(\beta^0)$, then the posterior distribution $\pi'(\alpha^0, \beta^0)$ can be fully characterized in closed-form as a bivariate normal distribution, i.e., $\pi'(\alpha^0, \beta^0)\sim \mathcal{N}\big(\pmb{\mu}, \pmb{\Sigma}\big)$, such that $\pmb{\mu} = (\mu_{\alpha^0}, \mu_{\beta^0})$ and $\pmb{\Sigma}=
\begin{bmatrix}
\delta^2_{\alpha^0} & r \\ 
r    &   \delta^2_{\beta^0}
\end{bmatrix}$. If we define the failure time $T^0$, as the first time ${A}^0(t)$ crosses a certain failure threshold $\Lambda$, then the nominal RUL distribution can be shown to follow an Inverse Gaussian distribution, i.e., $\lambda^0\sim \mathcal{N}^{-1}\bigg(\frac{\Lambda-d_{t^c}}{\mu_{\beta^0}},\big(\frac{\Lambda-d_{t^c}}{\sigma}\big)^2\bigg)$, where {$d_t^c$ is the observed degradation at the observation time $t^c$. Proof and discussion about this class of degradation models are found in \cite{yildirim2017integrated}.
 
\subsection{Time transformation model}
The baseline model in Section II.A. assumes that the WT ``always'' experiences nominal loading. In reality, WT blades are subject to time-varying loads largely due to the variations in wind speeds and yaw decisions. We assume that those time-varying loads directly impact the rate and diffusion of the degradation process, namely $\beta^0$ and $\sigma$. As such, we model the ``actual'' rate and diffusion of the degradation process as $\beta^0 \Psi(\phi(t;\tilde{\boldsymbol{\gamma}},\nu))$ and $\sigma\left[\Psi(\phi(t;\tilde{\boldsymbol{\gamma}},\nu))\right]^{\frac{1}{2}}$, respectively.
Here, $\Psi(\cdot)$ is a parametric function that expresses how higher loads (due to deviations from baseline conditions) accelerate (or decelerate) the degradation process. Specifically, in the baseline case, we have {$\Psi(\phi(t; \tilde{\gamma}^0,\nu^0)) = 1$.} Conversely, aggressive loading means $\Psi(\phi(t; \tilde{\boldsymbol{\gamma}}, \nu)) > 1$, implying faster-than-nominal degradation (i.e. the WT is expected to fail sooner), whereas  {{$\Psi(\phi(t; \tilde{\boldsymbol{\boldsymbol{\gamma}}}, \nu)) < 1$}} 
implies slower-than-nominal degradation process (i.e., longer expected RULs). 

In light of that, the actual amplitude of the degradation signal at time $t$, denoted by ${A}(t)$, is formulated as:
\begin{multline} \label{eq:degradation_function4}
    {A}(t)=\alpha^0+\int_0^t \beta^0\cdot \Psi\big(\phi(z;\tilde{\boldsymbol{\gamma}},\nu)\big)dz \\
    +\int_0^t\sigma\cdot\left[\Psi\big(\phi(z;\tilde{\boldsymbol{\gamma}},\nu)\big)\right]^{\frac{1}{2}}dW(z).
\end{multline}

\begin{lemma}
For the degradation model defined in (\ref{eq:degradation_function4}), the corresponding RUL, denoted by $\lambda$, has an inverse Gaussian distribution, i.e., $\lambda \sim \mathcal{N}^{-1}\big(\tau(t)|a^0,b^0\big), t\geq0$, where  $\tau(t)=\int_{0}^{t} \Psi(\phi(z;\tilde{\boldsymbol{\gamma}},\nu))dz$, $a^0=\frac{\Lambda-d_{t^c}}{\mu_{\beta^0}}$, and $b^0=\big(\frac{\Lambda-d_{t^c}}{\sigma}\big)^2$. \end{lemma}

The proof of Lemma 1 is provided in Appendix A. To elucidate the essence of Lemma 1 and how it acts as a ``time transformation,'' we provide a simple example with three cases. In the first case, let us consider a WT operating in baseline conditions, i.e. {{$\tilde{\boldsymbol{\gamma}}^0, \nu^0$}}. Hence, we would have 
{{$\Psi(\phi(t;\tilde{\gamma}^0, \nu^0))= 1$}}.
The second case considers higher-than-baseline loading conditions $\tilde{\gamma}^h, \nu^h$, such that $\Psi(\phi(t;\tilde{\gamma}^h, \nu^h))= 2$. The third case entails lower-than-baseline loading conditions $\tilde{\gamma}^l, \nu^l$, resulting in $\Psi(\phi(t;\tilde{\gamma}^l, \nu^l))= \frac{1}{2}$. The result in Lemma 1 entails mapping the calendar time $t$ into a transformed time $\tau(t)$, wherein the mapping is governed by the parametric function $\Psi(\cdot)$. For the above example, the values of $\Psi(\cdot)$ imply that two days of operation in baseline conditions is equivalent to one day of operation in high loading conditions, and four days of operation in low loading conditions, that is, we have $2 \Psi(\phi(t;\tilde{\gamma}^0, \nu^0))= \Psi(\phi(t;\tilde{\gamma}^h, \nu^h)) = 4\Psi(\phi(t;\tilde{\gamma}^l, \nu^l))$.

A key aspect of this time transformation idea is to specify $\Psi(\phi(t; \tilde{\gamma}, \nu))$ so that it best represents the degradation process given the loading up to time $t$, i.e., $\{\phi(z;\Tilde{\boldsymbol{\gamma}},\nu):0\leq z\leq t\}$. To do so, we follow a two-step procedure: In the first step ($\mathcal{S}1$), we find the impact of $\nu$ and $\tilde{\boldsymbol{\gamma}}$ on the structural loads,
i.e., $\{\tilde{\boldsymbol{\gamma}}, \nu\} \rightarrow \phi(t;\tilde{\boldsymbol{\gamma}}, \nu)$. In the second step ($\mathcal{S}2$), we find the impact of {loading}
on the degradation, i.e., $\phi(t;\tilde{\boldsymbol{\gamma}}, \nu) \rightarrow \Psi(\phi(t;\tilde{\boldsymbol{\gamma}}, \nu))$. For $\mathcal{S}1$, we directly use the data from \cite{kragh2014load} where the the load variation, namely the values of the function
{$\phi(.)$}, are provided at various inputs of $\nu$ and $\tilde{\boldsymbol{\gamma}}$. 

The second step, $\mathcal{S}2$, is less straightforward. Here, we make use of the $S$-$N$ curves, which relate the observed stress level to the number of cycles to failure, $N$, as expressed in (\ref{eq:sn_curve}).
\begin{equation} \label{eq:sn_curve}
   N=\bigg[\frac{1}{C_1}S(t;\tilde{\boldsymbol{\gamma}},{\nu})\bigg]^{-C_2} \sim \bigg[\frac{1}{C_1}\phi(t;\tilde{\boldsymbol{\gamma}},{\nu}) \bigg]^{-C_2},
\end{equation}
where $C_1$ and $C_2$ are the effective single cycle strength and fatigue strength exponent factors of the blades, respectively. 
The value of $C_2$ depends on the blade material. Using the $S$-$N$ curve relationship in (\ref{eq:sn_curve}) and the conclusions of Lemma 1, we can evaluate the impact of loading variations on the number of cycles, relative to those obtained under nominal conditions, as described in
(\ref{eq:degradation_function5}), which finds the number of cycles added or removed, relative to the nominal case.
{\begin{equation} \label{eq:degradation_function5}
   \varphi(t;\tilde{\boldsymbol{\gamma}},\nu): = \frac{N}{N^0}=\bigg[\frac{\phi(t;\tilde{\boldsymbol{\gamma}},\nu)}{\phi(t;\tilde{\boldsymbol{\gamma}}^0,\nu^0)}\bigg]^{-C_2},   
\end{equation}}
where $\phi(t;\tilde{\boldsymbol{\gamma}}^0,\nu^0)$ denotes the baseline load variations, while $N^0$ is the corresponding number of cycles to failure under nominal conditions.
Thus, {}
$\varphi(.)$ denotes the relative change in the number of cycles until failure due to deviations in wind speed and yaw misalignment relative to the nominal conditions, which directly translates to the relative gain/loss in RUL. That is, if a WT has a nominal RUL, $\lambda^0$, then, its ``actual'' RUL is predicted at time $t$ as in (\ref{eq:rul_final}). 

{\begin{equation}
\label{eq:rul_final}
\small
\underbrace{\lambda}_{\textit{Actual RUL}}=\underbrace{\lambda^0}_{\textit{Baseline RUL}}+ \underbrace{\int_{t^c}^{t}\bigg(1-\frac{1}{\varphi(z;\tilde{\gamma},\nu)}\bigg)dz}_{\textit{RUL gain/loss due to changes in $v$ and $\tilde{\boldsymbol{\gamma}}$}}.
\end{equation}}

Letting $f(\cdot) = \frac{1}{\varphi(\cdot)}$, and approximating the integral in (\ref{eq:rul_final}) by a sum yields a tractable expression for the actual, yaw-dependent RUL prediction, as shown in (\ref{eq:rul_final2}).
{\begin{equation}
\label{eq:rul_final2}
\lambda=
\lambda^0+\sum_{z=t^c}^{t}(1-f(z;\tilde{\gamma},\nu)).
\end{equation}}

\section{Optimization Model} \label{S:Opt_model}
POSYDON is formulated as a stochastic MILP, with a scenario set denoted by $\mathcal{S}$. We assume that an OSW farm consists of $N_{\mathcal{I}}$ WTs, with $i \in \{1, ..., N_{\mathcal{I}}\}$ denoting the WT's index. POSYDON considers two planning horizons: a day-ahead short-term horizon (STH) with hourly index $t\in\mathcal{T}$, 
and a long-term horizon (LTH), with daily index $d\in\mathcal{D}$, 
up to $N_{\mathcal{D}}$ days. The separation into short- and long-term horizons aligns with the standard practice in wind farm O\&M scheduling \cite{koltsidopoulos2020data}. 

\subsection{Decision Variables}
The decision variables in POSYDON are categorized into production and maintenance decisions, as detailed below: 

\subsubsection{Production Decisions}
\begin{itemize}
\item $\gamma_{t,i,j} \in \{0,1\}$ denotes whether the $j$th yaw misalignment (YM) level is selected at time $t$ for the $i$th WT. 
\item $\gamma^{L}_{d,i,j,s} \in \{0,1\}$ denotes whether the $j$th YM level is selected at day $d$ for the $i$th WT and $s$th scenario.
\end{itemize}
\subsubsection{Maintenance Decisions}
\begin{itemize}
\item $m_{t,i} \in \{0,1\}$ denotes whether the $i$th WT is selected for maintenance at time $t$.  
\item $m^{L}_{d,i,s} \in \{0,1\}$ denotes whether the $i$th WT is selected for maintenance at day $d$ under the $s$th scenario.
\end{itemize}

Two key assumptions are: (\textit{i}) we discretize the YM (originally a continuous variable) into discrete levels, $j \in \{1, ..., J\}$ for mathematical tractability. That is, $\boldsymbol{\gamma}$ (the discrete decision variable) denotes the level of YM to which $\tilde{\boldsymbol{\gamma}}$ (the continuous YM) belongs. To that end, we impose the constraints in (\ref{eq:gamma1}) and (\ref{eq:gamma2}) to ensure that at most one YM level is selected at a time; and (\textit{ii}) In practice, yaw updates are made at finer time intervals based on local wind direction variations \cite{scholbrock2014field}. Our yaw decisions can thus be considered as recommendations for the average YM level during the corresponding time periods.
\begin{equation}\label{eq:gamma1}
    \sum_{j\in \mathcal{J}} \gamma_{t,i,j} \leq 1 \quad \forall t \in \mathcal{T}, i \in \mathcal{I},
\end{equation}
\begin{equation}\label{eq:gamma2}
    \sum_{j\in \mathcal{J}} \gamma_{d,i,j,s}^L \leq 1 \quad \forall d \in \mathcal{D}, i \in \mathcal{I}, s \in \mathcal{S}.
\end{equation}

\subsection{Embedding the RUL Predictions in the Optimization}
The yaw-dependent RUL predictions obtained in (\ref{eq:rul_final2}) are embedded into the optimization model through (\ref{eq:yaw_degradation}), 
wherein the RUL (in days) of the $i$th WT and $s$th scenario, denoted by $\lambda_{i,s}$ is equal to the nominal RUL, $\lambda_{i,s}^{0}$, adjusted by the RUL gain/loss due to changes in YM decisions and the wind speed. The parameter $F_{t,i,j,s}$ denotes the relative RUL change from the baseline conditions, and represents a realization of the function $f(t; \tilde{\boldsymbol{\gamma}}, \nu)$ at time $t$, for the $i$th WT, under the $j$th YM decision, and $s$th scenario\textemdash Recall (\ref{eq:rul_final2}).

The binary parameter $\zeta_{i,s}^0$ denotes the operational status of the $i$th WT in the STH assuming nominal conditions. In specific, $\zeta_{i,s}^0=1$ when at least one day is left in the $i$th WT's nominal RUL estimate, i.e., when 
$\lambda_{i,s}^{0}\geq 1$. The same holds for the LTH, i.e., $\zeta_{d,i,s}^{0,L}=1~ \textit{iff}~ \lambda_{i,s}^{0}\geq d$.
\begin{multline} \label{eq:yaw_degradation}
\lambda_{i,s} = \lambda_{i,s}^{0} + \overbrace{\frac{\zeta_{i,s}^0}{24}\cdot\sum_{t\in \mathcal{T}}\left ( 1 - \sum_{j\in \mathcal{J}}\gamma_{t,i,j} \cdot F_{t,i,j,s} \right )}^{\textit{Day-ahead equivalent RUL gain/loss}}\\ + \underbrace{\sum_{d \in \mathcal{D}} \left [\zeta_{d,i,s}^{0,L} \cdot \left (1- \sum_{j\in \mathcal{J}} \gamma_{d,i,j,s}^L \cdot F_{d,i,j,s}^L \right) \right ]}_{\textit{LTH equivalent RUL gain/loss}} \quad  \forall i \in \mathcal{I}, s\in \mathcal{S}.
\end{multline}

Similarly, binary variables $\zeta_{i,s}$ and $\zeta_{d,i,s}^L$ denote the operational status of the $i$th WT under the actual conditions, in the STH and the LTH, respectively, as defined in (\ref{eq:zeta1})-(\ref{eq:zeta2L}).
\begin{equation} \label{eq:zeta1}
\zeta_{i,s} \leq \lambda_{i,s}, \quad  \forall i \in \mathcal{I}, s\in \mathcal{S},
\end{equation}
\begin{equation} \label{eq:zeta2}
\lambda_{i,s} - 1 \leq \text{M} \cdot \zeta_{i,s} , \quad  \forall i \in \mathcal{I}, s\in \mathcal{S},
\end{equation}
\begin{equation} \label{eq:zeta1L}
d \cdot \zeta_{d,i,s}^L \leq \lambda_{i,s}, \quad  \forall d \in \mathcal{D}, i \in \mathcal{I}, s\in \mathcal{S},
\end{equation}
\begin{equation} \label{eq:zeta2L}
\lambda_{i,s} - d \leq \text{M} \cdot \zeta_{d,i,s}^L , \quad  \forall d \in \mathcal{D}, i \in \mathcal{I}, s\in \mathcal{S},
\end{equation}
where M is an arbitrary large number. 
In (\ref{eq:zeta1L}), 
if $\lambda_{i,s}$, is less than $d$ days in scenario $s$, then $\zeta_{d,i,s}^L$ is forced to zero (i.e., the turbine fails at day $d$). Otherwise, $\zeta_{d,i,s}^L = 1$, as enforced by (\ref{eq:zeta2L}). The same holds for the STH through (\ref{eq:zeta1}) and (\ref{eq:zeta2}).

\subsection{Objective Function}
The objective function of POSYDON is shown in (\ref{eq:obj_fun}), and consists of four terms: the day-ahead profit, the long-term profit, the expected cost of prolonged interruptions, and the expected end-of-horizon cost. In the following sub-sections (\textit{T1}-\textit{T4}), we describe each of those terms separately. 
\begin{multline}\label{eq:obj_fun}
    \max_{\boldsymbol{m}, \pmb{\gamma}, \boldsymbol{m}^L, \pmb{\gamma}^L}
    \Bigg\{\overbrace{l^{STH}}^{\text{short-term profit}} + \overbrace{\sum_{d\in \mathcal{D}} l^{LTH}_d}^{\text{long-term profit}} \\ 
    - \frac{1}{N_S}\cdot\sum_{s\in\mathcal{S}} \sum_{i\in \mathcal{I}}\bigg [ \overbrace{\mbox{U}_s\cdot w_{i,s} + \mbox{Y}_s\cdot b_{i,s}}^{\text{prolonged interruptions}}\\ \underbrace{\small{-C^{\lambda}\cdot \bigg (\overbrace{\lambda_{i,s}}^{\text{RUL gain}} -\overbrace{\lambda_{i,s}^0\cdot\sum_{t\in\mathcal{T}}m_{t,i}- \sum_{d\in \mathcal{D}}(\lambda_{i,s}^0-d)\cdot m_{d,i,s}^L}^{\text{cycle days lost due to early maintenance}}  \bigg)} }_{\text{end of horizon cost}} \bigg ] \Bigg\}
\end{multline}

\subsubsection*{(T1) Short-term profit, $l^{STH}$} The short-term profit is defined in (\ref{STH_prof}) as the difference between the day-ahead operating revenue and the maintenance costs. The revenue is calculated as the product of the generated power, $p_{t,i,s}$, and the hourly electricity market price $\kappa_{t,s}$. The maintenance costs comprise four components: \textit{(i)} crew cost charged at the crew hourly rate $C^x$
, where $x_{t,i,s} \in \{0,1\}$ denotes whether a crew is dispatched to the $i$th WT at time $t$ for the $s$th scenario; \textit{(ii)} overtime cost charged at a rate of $C^q$
, where $q_s \in \mathbb{Z}^+$ denotes the overtime hours worked in the STH; and \textit{(iii)} daily vessel rentals at a daily rate of $C^r$
; and \textit{(iv)} repair cost (explained below). \begin{multline} \label{STH_prof}
    l^{STH}=  \frac{1}{N_S}\sum_{s \in \mathcal{S}}\bigg[\sum_{i \in \mathcal{I}} \sum_{t \in \mathcal{T}}(\overbrace{\kappa_{t,s} \cdot  p_{t,i,s}}^{\text{operating revenue}}   - \overbrace{C^x \cdot x_{t,i,s}}^{\text{crew cost}})\\ - \underbrace{C^q \cdot q_s}_{\text{overtime cost}}\bigg ] - \sum_{i \in \mathcal{I}} \sum_{t \in \mathcal{T}}\underbrace{(1-\rho_i)\cdot \xi_i \cdot  \alpha_{t,i}^m }_{\text{repair cost}} - \underbrace{C^r \cdot r}_{\text{vessel cost}}.
\end{multline} 

The repair cost entails the parameters $\rho_i \in \{0,1\}$ and $\xi_i\in\mathbb{R}^+$. The former denotes whether a maintenance task has already been initiated in a previous day but is yet to be completed. In this case, repair costs are deactivated, as they have already been paid for in a previous day. The paramater $\xi_i$ is a maintenance criticality coefficient which adjusts the weight of the maintenance cost. The variable $\alpha_{t,i}^m\in \mathbb{R}^+$ linearizes the product of $c_i \cdot m_{t,i}$, 
and is controlled by the additional set of linear constraints (\ref{prof_lin_1})-(\ref{prof_lin_3}), where variable $c_i \in \mathbb{R}^+$ is the ``dynamic maintenance cost (DMC)'' rate, which balances the trade-off between early and late maintenance actions, and is an explicit 
function of the RUL, $\lambda_{i,s}$, of the WT asset. 
Derivation and discussion on $c_i$ is deferred to Appendix B.    
\begin{equation}\label{prof_lin_1}
\alpha_{t,i}^{m} \leq  \mbox{M} \cdot m_{t,i} \quad  \forall t \in \mathcal{T}, i \in \mathcal{I}
\end{equation}
\begin{equation}\label{prof_lin_2}
\alpha_{t,i}^{m} \leq c_i \quad  \forall  t \in \mathcal{T}, i \in \mathcal{I}
\end{equation}
\begin{equation}\label{prof_lin_3}
\alpha_{t,i}^{m} \geq c_i  - \mbox{M}\cdot(1-m_{t,i}) \quad  \forall t \in \mathcal{T}, i \in \mathcal{I}
\end{equation}

\subsubsection*{(T2) Long-term profit, $l^{LTH}$} The LTH profit in (\ref{LTH_prof}) is similar to the STH profit, except for the crew work hours which are calculated as the expected mission time, $B_{d,i,s}^L$, which is the projected duration of a maintenance task considering both the repair time and the WT access conditions. Estimation of $B_{d,i,s}^L$ from actual wind/wave data is detailed in 
\cite{papadopoulos2022stochos}. 
\begin{multline} \label{LTH_prof}
l^{LTH}_d = \frac{1}{N_S}\sum_{s\in\mathcal{S}}\Bigg\{\sum_{i \in \mathcal{I}} \bigg[ \overbrace{\kappa_{d,s}^L\cdot p_{d,i,s}^L}^{\text{operating revenue}} - \overbrace{(1-\rho_i)\cdot\xi_i\cdot\alpha_{d,i,s}^{m,L}}^{\text{repair cost}}\\ - \underbrace{C^x\cdot B_{d,i,s}^L \cdot m_{d,i,s}^L}_{\text{crew cost}} \bigg] - \underbrace{C^r \cdot r_{d,s}^L}_{\text{vessel cost}} - \underbrace{C^q \cdot q_{d,s}^L}_{\text{overtime cost}}\Bigg\} \quad {\forall ~ d \in \mathcal{D}}
\end{multline}
Similar to the STH profit, $\alpha_{d,i,s}^{m, L} \in \mathbb{R}^+$, serves as an alias for the non-linear product $c_{d,i} \cdot m_{d,i,s}^L$, controlled by (\ref{prof_lin_4})-(\ref{prof_lin_6}), where $c_{d,i}$ is the LTH DMC rate\textemdash Refer to Appendix B.
\begin{equation}\label{prof_lin_4}
\alpha_{d,i,s}^{m,L} \leq \mbox{M} \cdot m_{d,i,s}^L \quad  \forall d \in \mathcal{D}, i \in \mathcal{I}, s\in \mathcal{S}
\end{equation}
\begin{equation}\label{prof_lin_5}
\alpha_{d,i,s}^{m,L} \leq c_{d,i}^L \quad  \forall  d \in \mathcal{D}, i \in \mathcal{I}, s\in \mathcal{S}
\end{equation}
\begin{equation}\label{prof_lin_6}
\alpha_{d,i,s}^{m,L} \geq c_{d,i}^L  - \mbox{M}\cdot(1-m_{d,i,s}^L) \quad  \forall d \in \mathcal{D}, i \in \mathcal{I}, s\in \mathcal{S}
\end{equation}

{\subsubsection*{(T3) Prolonged Interruptions} This term captures the cost associated with the maintenance tasks that may be initiated in the STH but would have to be interrupted and (later) completed in the LTH due to unfavorable weather conditions. We let $w_{i,s} \in \{0,1\}$ denote the occurrence of such event, such that $\mbox{U}_s \cdot w_{i,s}$ represents the corresponding upfront costs (e.g. extra vessel rental costs). The variable $b_{i,s}\in\mathbb{Z}^+$ is the remaining time to complete the unfinished maintenance task in the LTH, while the parameter $\mbox{Y}_s$ denotes the cost of additional hourly revenue losses until the maintenance is completed.}  

\subsubsection*{(T4) End of Horizon Cost} This term encodes the benefit of prolonging the RUL of the WTs beyond the optimization horizon (that is, beyond the LTH), in order to discourage the optimization from unnecessarily scheduling premature preventive maintenance tasks within the foreseeable future, especially if the WT has plenty of buffer in its RUL prediction. This is primarily controlled by the parameter $C^{\lambda}~(\$/\mbox{day})$ which approximates the economic gain/loss per day of RUL. 

\subsection{Other Constraints}
The remaining constraints, used to model different aspects of O\&M in an OSW farm, are listed as follows.

\subsubsection*{Maintenance constraints} The equality in (\ref{eq:maints}) forces a maintenance to be scheduled either in the STH or the LTH.
\begin{equation} \label{eq:maints}
\sum_{t \in \mathcal{T}}m_{t,i}+\sum_{d \in \mathcal{D}}m_{d,i,s}^L = \theta_i \quad \forall  ~ i \in \mathcal{I}, s \in \mathcal{S},
\end{equation}
where $\theta_i \in \{0,1\}$ denotes whether the $i$th WT requires maintenance in the near future, and is determined as in (\ref{eq:theta}).
\begin{equation} \label{eq:theta}
N^{\theta} - \lambda_{i,s} \leq \mbox{M}\cdot \theta_i \quad \forall  ~ i \in \mathcal{I}, s \in \mathcal{S},
\end{equation}
such that $N^{\theta}$ is 
parameter denoting the lower time threshold for the RUL scenarios, below which, the $i$th WT has to be scheduled for maintenance in the optimization horizon.

Once a maintenance task is initiated at time $t$ for the $i$th WT, then it would be under maintenance for a period of time computed as the minimum between the remaining time in the STH, which is $24-t$, and the mission time $B_{t,i,s}$. This is expressed in (\ref{eq:downs}), where $u_{\tilde{t},i,s} = 1$ denotes a WT under maintenance at time $\tilde{t}$.  
\begin{multline} \label{eq:downs}
\sum_{\tilde{t}=t}^{\min\{24, t+B_{t,i,s}\}} u_{\tilde{t},i,s} \geq \min\{24-t,B_{t,i,s}\} \cdot m_{t,i} \\\quad \forall t \in \mathcal{T}, i \in \mathcal{I}, s\in \mathcal{S}.
\end{multline}

If that maintenance task is not completed within the STH, then $b_{i,s} \in \mathbb{Z}^+$, as shown in (\ref{eq:incomplete2s}), denotes the remaining time required to complete it in the LTH, while $w_{i,s} \in \{0,1\}$, as shown in (\ref{eq:incompletes}), denotes the occurrence of such event and is only set to $1$ once $b_{i,s} > 0$. This is the case where the ``prolonged interruption'' term in (\ref{eq:obj_fun}) is activated.
\begin{equation} \label{eq:incomplete2s}
{b_{i,s} \geq \sum_{t\in \mathcal{T}}m_{t,i}\cdot [B_{t,i,s}- 24 +t ]^+
 \quad \forall i \in \mathcal{I}, s\in \mathcal{S}},
\end{equation}
\begin{equation} \label{eq:incompletes}
{b_{i,s} \leq \mbox{M} \cdot w_{i,s}
 \quad \forall i \in \mathcal{I}, s\in \mathcal{S}}.
\end{equation}

The maintenance crew, once dispatched, is occupied until the maintenance is completed or their shift ends by the time of last sunlight, $t_D$, as shown in (\ref{eq:crews}). An upper bound on the number of crews is set to $N^x$ (crews), as shown in (\ref{eq:crew2s}).
\begin{equation} \label{eq:crews}
{x_{t,i,s} \geq u_{t,i,s} - \frac{t}{t_D}
\quad \forall t \in \mathcal{T}, i \in \mathcal{I}, s\in \mathcal{S}},
\end{equation}
\begin{equation} \label{eq:crew2s}
\sum_{i \in \mathcal{I}} x_{t,i,s} \leq N^x \quad \forall t \in \mathcal{T}, s\in \mathcal{S}.
\end{equation}

\subsubsection*{Turbine availability and power constraints}
A failed WT (i.e., one which has not been maintained at or before its RUL), or a WT currently undergoing maintenance, remains unavailable until a maintenance action is completed. In case of the STH, this can be expressed as in (\ref{eq:availabilitys}).
\begin{multline} \label{eq:availabilitys}
y_{t,i,s} \leq \overbrace{\zeta_{i,s}\cdot(1-\rho_i)}^{\text{WT operational status}} + \overbrace{\frac{24 \cdot \sum_{\tilde{t} \in \mathcal{T}}m_{\tilde{t},i}- \sum_{\tilde{t} \in \mathcal{T}}(\tilde{t} \cdot m_{\tilde{t},i})}{24-t}}^{\text{availability restored after maintenance}} \\
\quad \forall t \in \mathcal{T}, i \in \mathcal{I}, s\in \mathcal{S}. 
\end{multline}
The first term of the right hand side in (\ref{eq:availabilitys}) denotes whether the turbine is in a failed state at the beginning of the STH, or if a maintenance task has been initiated in a previous day. 
In case $\lambda_{i,s}<1$, then the turbine fails ($\zeta_i=0$), and can only return to its former operational status once a CM action is performed, guaranteed by the second term in the right hand side.
A similar constraint for the LTH is shown in~(\ref{eq:availability2s}).
\begin{multline} \label{eq:availability2s}
y_{d,i,s}^L \leq {\zeta_{d,i,s}^L\cdot (1-\rho_i)}+\frac{N_{\mathcal{D}}- \sum_{\tilde{d} \in \mathcal{D}}(\tilde{d} \cdot m_{\tilde{d},i,s}^L)}{N_{\mathcal{D}}-d} \hspace{0.3cm} \\\forall d \in \mathcal{D}, i \in \mathcal{I}, s\in \mathcal{S}. 
\end{multline}
A WT under maintenance remains unavailable until the task is completed, which is ensured by (\ref{eq:maintts}).
\begin{equation} \label{eq:maintts}
   y_{t,i,s} \leq 1 - u_{t,i,s} \quad \forall t \in \mathcal{T}, i \in \mathcal{I}, s\in \mathcal{S}.
\end{equation}

To compute the power output, we utilize the {yaw-adjusted additive multivariate kernel (YAMK) method proposed in \cite{pranjaal2022yaw}. The YAMK approach extends kernel-based multivariate power curve models \cite{golparvar2021surrogate, ding2019data} by integrating YM as an exogenous input via a local polynomial regression formulation. The output of YAMK is the expected scaled power output, $f_{t,i,j,s}$, as function of wind speed and YM.} This is expressed in (\ref{eq:yaw_power_STH}), where $\mbox{R}$ denotes the WT rated capacity (in MW). \begin{equation}\label{eq:yaw_power_STH}
    p_{t,i,s}\leq\mbox{R} \cdot \sum_{j\in \mathcal{J}} (\gamma_{t,i,j} \cdot f_{t,i,j,s}) \quad \forall t \in \mathcal{T}, i \in \mathcal{I}, s\in \mathcal{S}.
\end{equation}
Likewise, the daily power output $p_{d,i,s}^L$ is defined in (\ref{eq:yaw_power_LTH}), where $f_{d,i,j,s}^L \in [0, 1]$ is the daily yaw-dependent scaled power. 
\begin{equation}\label{eq:yaw_power_LTH}
    p_{d,i,s}^L\leq 24\cdot \mbox{R} \cdot \sum_{j\in \mathcal{J}} (\gamma_{d,i,j,s}^L \cdot f_{d,i,j,s}^L) \quad \forall d \in \mathcal{D}, i \in \mathcal{I}, s\in \mathcal{S}.
\end{equation}
The power output of the WT is constrained by its availability. For the STH, this is expressed by (\ref{eq:eqnoms}).
\begin{equation}\label{eq:eqnoms}
    p_{t,i,s}\leq {\mbox{R}} \cdot y_{t,i,s} \quad \forall t \in \mathcal{T}, i \in \mathcal{I}, s\in \mathcal{S}.
\end{equation}
For the LTH, this is expressed by (\ref{eq:eqnomsL}).
\begin{equation} \label{eq:eqnomsL}
p_{d,i,s}^L\leq 24 \cdot \mbox{R} \cdot  y_{d,i,s}^L \quad \forall d \in \mathcal{D}, i \in \mathcal{I}, s\in \mathcal{S}.
\end{equation}
To account for the impact of PM actions on the power output in the LTH, we use (\ref{eq:mainttsL}) as an approximation, where $f_{d,i,s}^{\max,L}$ is the maximum scaled power output given the wind conditions.
\begin{equation} \label{eq:mainttsL}
p_{d,i,s}^L\leq \mbox{R} \cdot f_{d,i,s}^{\max,L} \cdot (24 -B_{d,i,s}^L \cdot  m_{d,i,s}^L) \quad \forall d \in \mathcal{D}, i \in \mathcal{I}, s\in \mathcal{S}.
\end{equation}

\subsubsection*{Resource constraints}
Vessels are rented daily only if a maintenance task has been scheduled at that day. This is enforced via (\ref{eq:vessel1s})-(\ref{eq:vessel2s}), for the STH and LTH, respectively.
\begin{equation} \label{eq:vessel1s}
    \mbox{M}\cdot r \geq \sum_{t \in \mathcal{T}} \sum_{i \in \mathcal{I}} m_{t,i},
\end{equation}
\begin{equation}\label{eq:vessel2s}
  \mbox{M} \cdot r_{d,s}^L \geq \sum_{i \in \mathcal{I}}m_{d,i,s}^L \quad \forall d \in \mathcal{D}, s\in \mathcal{S}.
\end{equation}

The total work hours of the maintenance crews, cannot exceed $N^q$ (hour/crew), as we show in (\ref{eq:workhourss}). 
Otherwise, overtime hours, tracked by variable $q_s\in \mathbb{R}^+$, are incurred and compensated at a higher rate determined by $C^q$ (\$/hour). An upper bound of $N^H$ limits the total number of overtime hours in the STH, as expressed in (\ref{eq:overtime1s}). 
\begin{equation} \label{eq:workhourss}
\sum_{t \in \mathcal{T}, i \in \mathcal{I}} x_{t,i,s} \leq N^x \cdot N^q+q_s \quad \forall s\in \mathcal{S},
\end{equation}
\begin{equation} \label{eq:overtime1s}
q_s \leq N^H \quad \forall s\in \mathcal{S}.
\end{equation}
A similar set of constraints is introduced for the LTH.
A task that has started but not finished in the STH is prioritized to be picked up again at the first day of the LTH ($d = 1$). This is expressed in (\ref{eq:workhours2s}) where $B_{d=1,i,s}^L \cdot m_{d=1,i,s}^L + b_{i,s}$ denotes the total work hours in the first day of the LTH (the sum of the mission times and the remaining maintenance time of unfinished tasks). A similar constraint in (\ref{eq:workhours3s}) is imposed for the remaining days of the LTH, i.e., for $d \in \{2, ..., N_D\}$. Finally, (\ref{eq:overtime2s}) limits the total overtime hours in the LTH.
\begin{equation} \label{eq:workhours2s}
\sum_{i \in \mathcal{I}} \left[B_{1,i,s}^L \cdot m_{1,i,s}^L + b_{i,s}\right] \leq N^x \cdot N^q + q_{1,s}^L \quad \forall 
s\in \mathcal{S},
\end{equation}
\begin{equation} \label{eq:workhours3s}
\sum_{i \in \mathcal{I}} \left[B_{d,i,s}^L \cdot m_{d,i,s}^L\right] \leq N^x \cdot N^q + q_{d,s}^L \quad \forall 
d \in \{2, ..., N_{\mathcal{D}}\}, s\in \mathcal{S},
\end{equation}
\begin{equation} \label{eq:overtime2s}
q_{d,s}^L \leq N^H \quad \forall d \in \mathcal{D}, s\in \mathcal{S}.
\end{equation}

\section{Numerical Experiments} \label{S:numerical_exp}

We start this section by a description of the data and experimental setup used to evaluate the effectiveness of our proposed approach.
We then present the results and discussion. 

\subsection{Data Description} \label{S:data}
We use real-world met-ocean data and numerical weather predictions (NWPs) from the U.S. NY/NJ Bight\textemdash a region where several large-scale OSW projects are currently in-development \cite{lease2021}. In specific, wind speed and wave height data are obtained from a recently deployed (and publicly available) measurement campaign by NYSERDA \cite{nyserda2019}. Co-located NWPs for wind speed and wave height are obtained from the RU-WRF \cite{optis2020validation} and WAVEWATCH III \cite{wavewatch} models, respectively. Electricity price data {are obtained from PJM Data Miner III \cite{PJM} at node COMED (node id: 33092371).
Forecasts for the same node are obtained by the LEAR method proposed in \cite{lago2021forecasting}.} The processed datasets have an hourly resolution, spanning from August 12, 2019 to April 29, 2020. Scenarios for all aforementioned variables are generated using a probabilistic forecasting method discussed in \cite{papadopoulos2022stochos} and based on a Gaussian Process framework \cite{ezzat2018spatio, ye2023airu}. The resulting scenarios (or trajectories) naturally encode the temporal dependencies of the uncertain input parameters.  

The power production parameters, namely  $f_{t,i,j,s}$ and $f_{d,i,j,s}^L$ are obtained by plugging in the wind speed and YM levels to the yaw-dependent power curve model obtained using the YAMK method proposed in \cite{pranjaal2022yaw}, as shown in Figure \ref{fig:yaw_power}. For the degradation signals, we use experimental vibration data from a rotating machinery, subjected to accelerated life testing, which are described in detail in \cite{gebraeel2005residual}.Simulated FBM load data as function of wind speed and YM are obtained from \cite{kragh2014load}.
The relative change in RUL that is used to characterize $F_{t,i,j,s}$ and $F_{d,i,j,s}^L$ is estimated using (\ref{eq:degradation_function5}), and is illustrated in Figure \ref{fig:yaw_RUL_loss}, wherein wind speed and YM combinations that alleviate the loading and, therefore, extend the RUL ($F<1$), are colored in blue, whereas the combinations that increase the loading, thereby decreasing the RUL ($F>1$), are colored in red. 
At nominal loading conditions ($\tilde{\gamma}^0=0^{\degree}, \nu^0=10$ ms$^{-1}$), the RUL is neither extended, nor reduced from the nominal case ($F=1$).
{
Regions with extreme RUL losses ($F>>1$) would translate to extreme loading levels, and consequently, immediate failure. Hence, those regions are naturally avoided by the optimization solver as it 
significantly inflates the maintenance costs.} 

\begin{figure}[]
    \centering
    \includegraphics[width=0.9\linewidth]{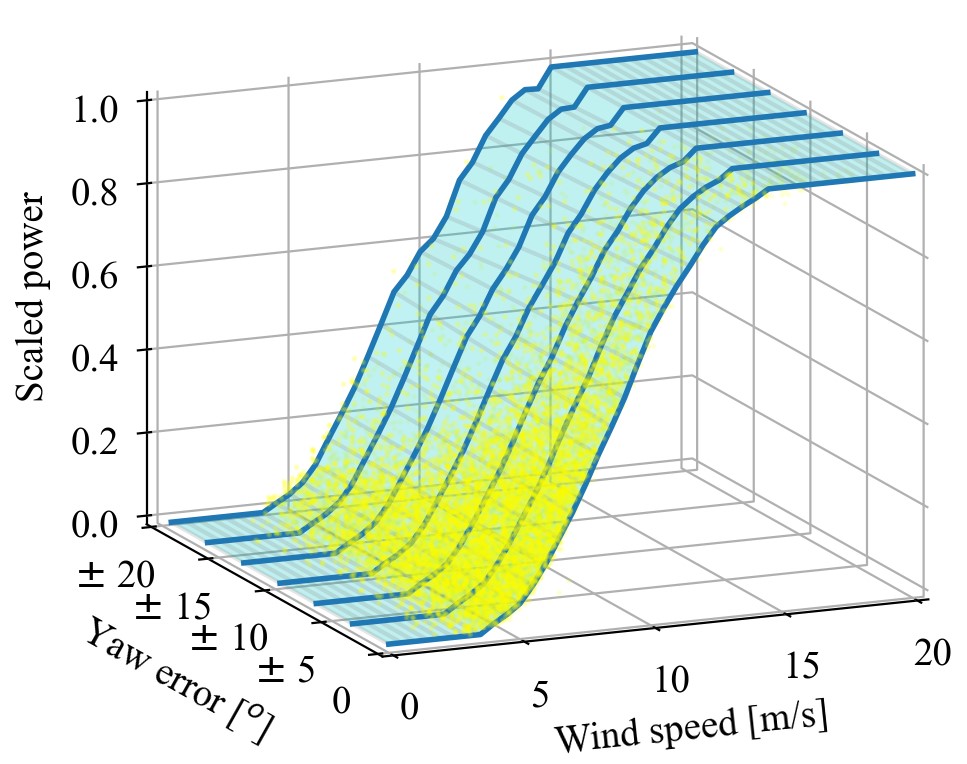}
    \caption{Yaw-dependent power curve obtained using the YAMK method \cite{pranjaal2022yaw}. Yellow points are the actual observations on which the power curve is trained; The cyan surface represents the hyperplane formed by the estimated power curve; Blue lines show the discrete power curves used by POSYDON.}
    \label{fig:yaw_power}
\end{figure}

\begin{figure}[]
    \centering
    \includegraphics[width=\linewidth]{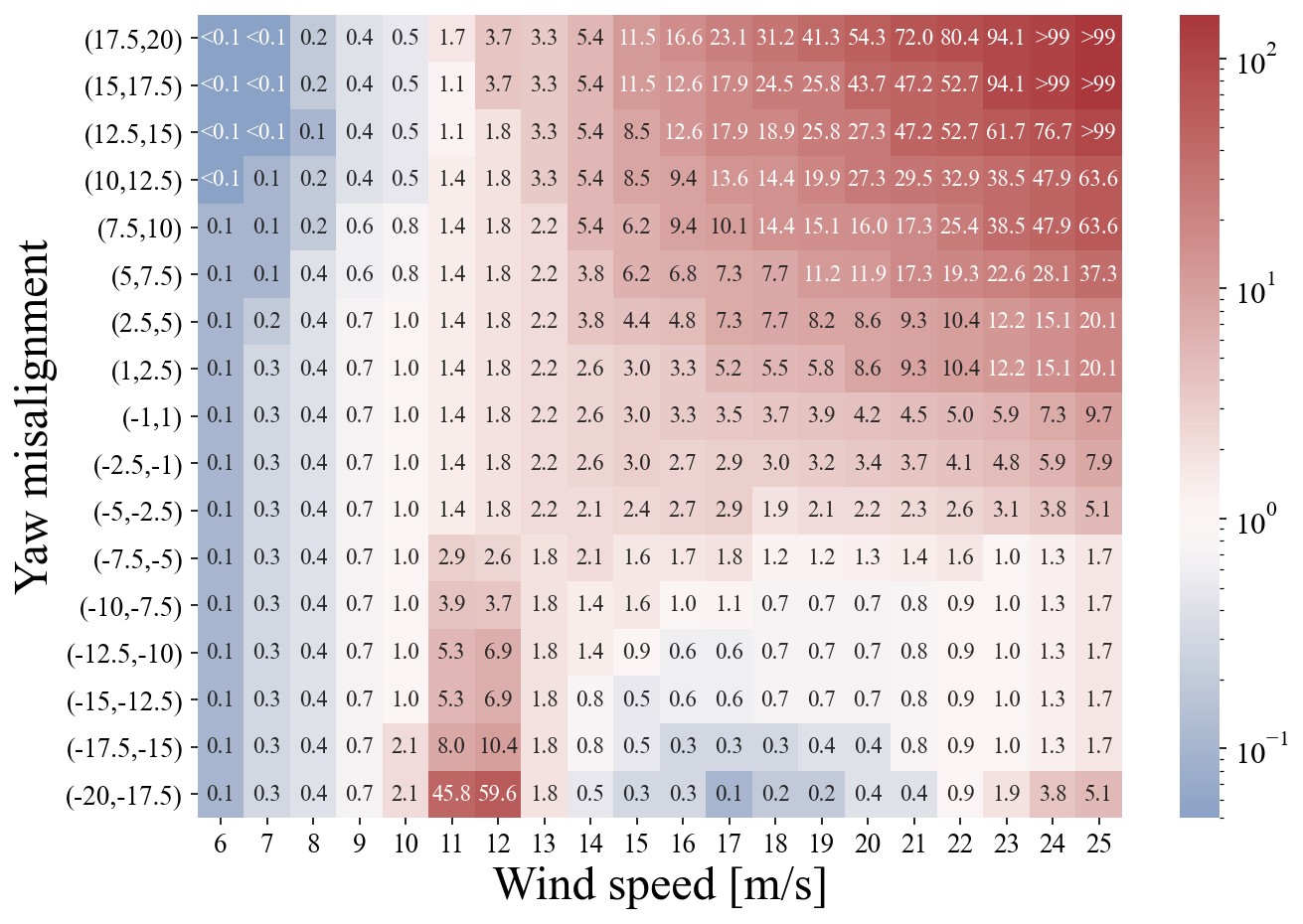}
    \caption{RUL change, relative to the nominal case, as a function of the wind speed and the YM decisions. This table was created by implementing our proposed approach in Section II on the loading data from \cite{kragh2014load}.}
    \label{fig:yaw_RUL_loss}
\end{figure}

\subsection{Experimental Setup}
Two sets of experiments are conducted. The first case study considers a $5$-WT OSW farm. This small test case allows us to extract meaningful ({and visual}) insights about the performance of POSYDON. The second case study considers a $50$-WT OSW farm, aiming to validate POSYDON in a realistic OSW farm setting. {Table \ref{tab:pars} shows the parameter values in our experiments, which were largely based on the experimental setup in \cite{papadopoulos2022stochos}. With the exception of $C^\lambda$, all parameter selections therein are made in light of credible prior studies on OSW O\&M practices\textemdash See \cite{papadopoulos2022stochos} for an elaborate discussion. The end of horizon cost coefficient, $C^\lambda$ is estimated as the daily weighted average of the repair costs reported in \cite{carroll2016failure}.} We set $N_{\mathcal{S}}=50$ scenarios and $N_{\mathcal{J}}=7$ YM levels of $5^{\degree}$ width, centered at 
$0^{\degree}$. A rolling optimization horizon of $10$ days is considered, i.e., $N_{\mathcal{D}}=9$ days, as the wind forecast quality largely reduces beyond that horizon \cite{ezzat2019spatio, ezzat2020turbine}. Once a solution is obtained, we slide forward by one day and re-optimize. We repeat the sliding process for 
$238$ days. 

POSYDON is compiled  in Python and solved using the Gurobi (v 9.1.2) solver for a relative optimality gap of $0.1\%$ and a time limit of $30$ minutes, on  
a server with two $14$-core Intel Xeon CPUs with a base frequency of $2.60$ GHz each, and total RAM of $128$ GB. The mean solution times for the $5$- and $50$-WT cases are $3.27$
and $20.26$ minutes, respectively. {A pseudocode detailing the implementation of POSYDON is presented in Appendix C.}

\begin{table}[]
   \caption{Parameter values used for the two sets of experiments}
    \renewcommand{\arraystretch}{1}
    \centering
    \begin{tabular}{ c c c}
      \hline
      Notation &    Parameter   &   Value   \\
      \hline
      $C^{\text{PM}}$   &  Preventive maintenance cost &   \$$4,000$\\
      $C^{\text{CM}}$   & Corrective maintenance cost    &   \$$10,000$\\
      $N^x$    &   Maximum \# of crews &   $2$ crews   \\
      $C^x$    &   Crew hourly rate    & \$$250$/hour  \\
      $C^q$    &   Overtime hourly rate  &   \$$125$/hour \\
      $N^q$    &   Maximum \# of regular work hours   &   $8$ hours   \\
      H     &   Maximum \# of overtime work hours & $8$ hours \\
      $C^r$  &   Daily vessel rental cost  &    \$$2,500$/day  \\
      $\nu_{max}$   &   Wind speed safety threshold &   $15$ m/s    \\
      $\eta_{max}$  &   Wave height safety threshold &   $1.8$ m \\
      $t_R$     &   Time of first light & 6:00 am  \\
      $t_D$ &   Time of last sunlight   & 9:00 pm  \\
      $C^{\lambda}$ &   End of horizon cost coefficient   & \$$30$/day  \\
      \hline
    \end{tabular}
     \label{tab:pars}

\end{table}

\subsection{Results}
The following benchmarks are used to evaluate POSYDON: (\textit{i}) \textbf{STOCHOS}: an adaptation of a state-of-the-art opportunistic maintenance optimization model in \cite{papadopoulos2022stochos}, where YM decisions are not considered, and hence are set to $0^{\degree}$. In this way, this benchmark would be equivalent to separately optimizing maintenance (using STOCHOS) and then maximizing production (irrespective of degradation) by always seeking minimal YM; (\textit{ii}) \textbf{DET}: the deterministic counterpart of POSYDON, primarily used to showcase the value of considering uncertainty in OSW operations; 
(\textit{iii}) \textbf{TBS}: Maintenance and production are optimized separately. For maintenance, we use a standard time-based (also called periodic) strategy with $60$-day intervals, and the YM decision is always set at $0^{\degree}$. 

\subsubsection{The $5$-WT Case Study}
Figure \ref{fig:base_schedule} depicts the degradation profiles and the maintenance schedules {for all competing models, namely: POSYDON, STOCHOS, DET, and TBS.} Looking at Figure \ref{fig:base_schedule}, few key insights can be drawn. 
First, we can immediately see how the maintenance cycles in the strategies that jointly optimize production and maintenance (POSYDON and DET) are relatively longer than those that do not (STOCHOS and TBS). This is because the joint optimization approach is able to tolerate\textemdash whenever economically justified\textemdash non-zero YM, thus partially sacrificing short-term power revenues for the more economically attractive target of prolonging the RUL, especially for critically degraded WTs. In contrast, the faster degradation in STOCHOS, incurred by always aggressively seeking for perfect YM, irrespective of the ensuing accumulating degradation, appears to be short-sighted, leading to more frequent maintenance requirements on the long-run. 
\begin{figure}[]
    \centering
\includegraphics[width=\linewidth]{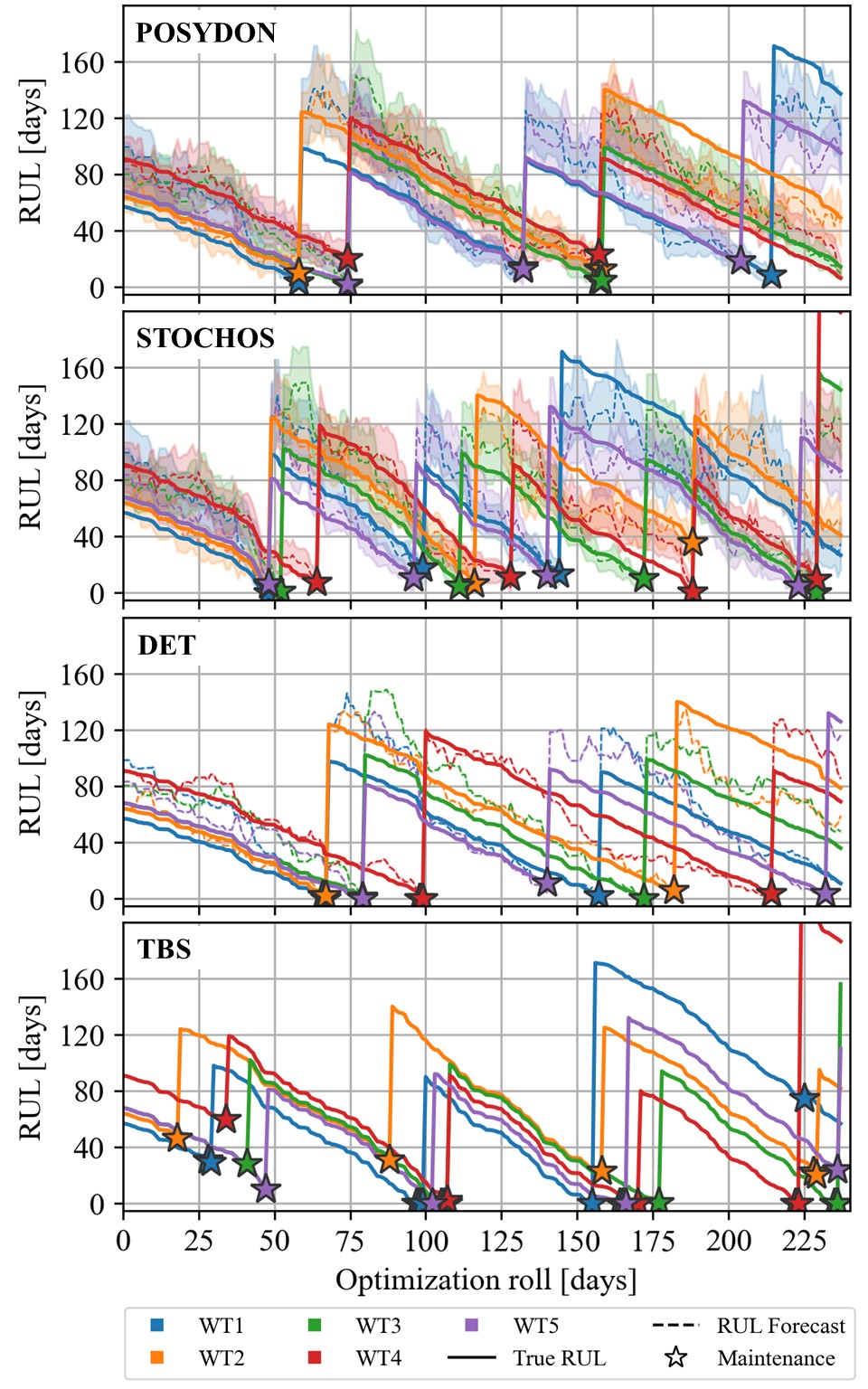}
    \caption{{The maintenance schedules for the $5$-WT case study using POSYDON, STOCHOS, DET, and TBS. Stars indicate an initiated maintenance task for the respective WT. Solid lines denote the true RUL profiles. For stochastic models (POSYDON and STOCHOS), shaded regions show the mean $\pm 1$ standard deviation of the RUL scenarios, whereas dashed lines in POSYDON, STOCHOS, and DET show the correspondent point RUL forecasts.}} 
    \label{fig:base_schedule}
\end{figure}

Second, by comparing the maintenance schedules of POSYDON and STOCHOS, we can see that the former groups maintenance tasks more effectively. This is despite both POSYDON and STOCHOS being, by design, `opportunistic'' strategies, i.e., they are both designed to incentivize the grouping of maintenance tasks when economically justified (e.g., to share maintenance setup costs or harness periods of low winds). The key distinguishing feature that enables POSYDON to achieve a more efficient grouping is its ability to loosely control the degradation process through YM decisions. 
In contrast to STOCHOS, POSYDON attempts to synchronise the degradation profiles for WTs sharing similar degradation levels, thereby achieving better grouping. However, to fully harness this ability, incorporating the forecast uncertainty in the decision-making process is key to hedge against the impact of forecast errors. That is evident from the superior grouping performance of POSYDON relative to DET.

\begin{table}[]
   \caption{Comparison of O\&M metrics for the $5$-WT case}
    \renewcommand{\arraystretch}{1}
    \centering
    \begin{tabular}{l | c c c c}
      \hline
      & \multicolumn{4}{c}{\textbf{O\&M Strategy}} \\
        \textbf{O\&M Metric} & {POSYDON} & {STOCHOS} & {DET} & {TBS} \\
      \hline
      Total Cost (K\$)  &  $\bf{159.3}$ & $273.8$ & $398.4$ & $715.0$ \\
      Revenue Koss (K\$) & $\bf{74.4}$ & $121.6$ & $283.5$ & $475.8$ \\
      Production Loss (GWh) & $\bf{1.5}$ & $2.5$ & $5.8$ & $9.7$ \\
      Maintenance Outages & ${12}$ & $18$ & $\bf{11}$ & $20$ \\
      Of which Corrective & $\bf{0}$ & $3$ & $4$ & $8$ \\
      \# of Vessel Rentals & $\bf{7}$ & $14$ & $11$ & $28$ \\
      \hline
    \end{tabular}
    \label{tab:exp1_results}
\end{table}

Table \ref{tab:exp1_results} shows the quantitative performance of all benchmarks in the $5$-WT case across several O\&M metrics, namely total costs, revenue losses, production losses, total number of maintenance tasks, number of corrective maintenance tasks, and number of vessel rentals. Importantly, POSYDON achieves significant cost improvements relative to its competitors, in particular 
$41.8\%$ reduction relative to STOCHOS, $60.0\%$ reduction relative to DET, {and $77.7$\% relative to TBS.} 

We also note that production losses are one of the primary determinants of POSYDON's superior performance. Compared to STOCHOS, POSYDON achieves $40.0\%$ reduction in production losses. This is especially interesting given that STOCHOS primarily employs a short-term production maximization strategy (seeking perfect YM). However, the long-term reductions in downtime due to prolonged RULs achieved by POSYDON significantly outweighs the benefits of aggressively harnessing short-term production gains. 
Over the span of the $238$ days, POSYDON requires $12$ maintenance tasks (none of which are corrective, i.e., post-failure), as opposed to the $15$ preventive and $3$ corrective maintenance tasks for STOCHOS, which translate to larger economic losses due to higher maintenance costs and frequent repair shutdowns. It is also worth noting that, although DET requires a slightly lower number of maintenance tasks ($11$ compared to $12$ for POSYDON), $4$ of those $11$ tasks are corrective, which are much costlier than preventive actions, and are primarily attributed to DET's inability to recognize the forecast uncertainties, and hence, having to react to unexpected failures.  

\subsubsection{The $50$-WT case}
Similar insights can be drawn from the second case study. Looking at Table \ref{tab:exp2_results}, we see that POSYDON leads to solutions which are more than $20\%$ better in terms of total cost, while yielding $17\%$ smaller production losses relative to STOCHOS (its closest competitor). Again, this is attributed to the increased maintenance requirements of STOCHOS, which ultimately inflates its overall O\&M costs. 

{The $50$-WT case study can provide further interesting insights on the interplay between the short- and long-term considerations. Here, we added three additional O\&M metrics that are more relevant to the $50$-WT case: (\textit{i}) The total downtime (in days); (\textit{ii}) the downtime incurred due to accessibility limitations (i.e., a WT that needs maintenance, yet is inaccessible due to harsh weather conditions); and (\textit{iii}) lost cycle days per task as a proxy for the cost of early maintenance (i.e., how long further the WT was expected to still operate had it not been shut down for maintenance). Analyzing those metrics, we observe that both stochastic models (i.e. POSYDON and STOCHOS) are the least impacted by access-related interruptions. Yet, with its ability to loosely control the WT degradation, POSYDON more effectively evades periods where maintenance must be performed during periods of unfavorable weather conditions, thus resulting in significantly lower access-related downtime. We also find that POSYDON yields the smallest number of lost cycle days per task, which practically means that the other strategies are more prone to committing pre-mature/early maintenance actions, thus unnecessarily incurring significant production losses.} 

{Interestingly, although DET has similar maintenance requirements to POSYDON in both case studies, the production loss it experiences is even higher than that of STOCHOS.
In addition to its risk-taking attitude which typically leads to an increased number of corrective tasks, a major drawback of DET lies in its inherent assumption of perfect information. 
We particularly observe that DET is highly impacted by wind forecasting errors for two reasons: (1) wrongly forecasting the WT access state (which is a function of the wind and wave forecasts), and hence having to delay the now-infeasible maintenance actions; and (2) the non-linear form of the power curve, which inflates the wind speed forecast errors when translated to the power domain. Combined, this leads DET to make decisions that over-prioritize the long-term maintenance expenses of the WT, which significantly compromises its shorter-term power production revenues.} 

Finally, we notice that TBS performs poorly across all O\&M metrics in both sets of experiments.
This is the result of its rigidness with respect to the WT's health condition, as well as 
its inability to opportunistically group maintenance actions, leading to an inflated number of vessel rentals with little utilization rate and high production losses.
Although TBS has the highest maintenance frequency, it results in a significantly higher number of failures, pointing to the poor performance of O\&M strategies that consider rigid maintenance intervals, {\textit{sans} sensory information}, especially in an OSW environment.

\begin{table}[]
   \caption{Comparison of O\&M metrics for the $50$-WT case}
    \renewcommand{\arraystretch}{1}
    \centering
    \begin{tabular}{l | c c c c}
      \hline
      & \multicolumn{4}{c}{\textbf{O\&M Strategy}} \\
        \textbf{O\&M Metric} & {POSYDON} & {STOCHOS} & {DET} & {TBS} \\
      \hline
      Total Cost (M\$)  &  $\bf{2.5}$ & $3.1$ & $3.6$ & $6.1$ \\
      Revenue Loss (M\$) & $\bf{1.6}$ & $1.9$ & $2.7$ & $4.4$ \\
      Production Loss (GWh) & $\bf{32.0}$ & $38.7$ & $54.9$ & $88.2$ \\
      {Downtime (days)} & $\bf{190.2}$ & $247.4$ & $328.8$ & $504.3$ \\
     {Access Downtime (days)} & $\bf{158.2}$ & $215.4$ & $296.8$ & $455.2$ \\
     {Lost Cycle Days/Task} & $\bf{7.8}$ & $9.7$ & $9.3$ & $22.8$ \\
      Maintenance Outages & $107$ & $155$ & $\bf{103}$ & $173$ \\
      Of which Corrective & $\bf{9}$ & $14$ & $11$ & $58$ \\
      \# of Vessel Rentals & $64$ & $86$ & $\bf{62}$ & $137$ \\
      
      \hline
    \end{tabular}
    \label{tab:exp2_results}
\end{table}

\section{Conclusions} \label{S:concl}
In this work, we proposed an {opportunistic} decision-theoretic framework to jointly optimize production and maintenance decisions in an OSW farm, and ultimately lower its O\&M expenses. Based on a stochastic MILP framework, POSYDON effectively links the impact of short-term production decisions on long-term maintenance expenses, and vice-versa. The merit of this co-optimization approach is illustrated through a set of real-world experiments, wherein POSYDON demonstrates significant improvements in terms of several O\&M metrics, relative to a set of prevalent benchmarks. 

{Beyond this work, we are currently investigating how to integrate additional decision dependencies into POSYDON, such as multi-turbine dependencies induced by wake interactions and spatial effects. We also do believe that the O\&M models and formulations proposed herein can be transferable to other forms of renewable energy generation that bear similarities with OSW. In addition to land-based wind farms, similar short- and long-term trade-offs exist in wave energy converters, where accessibility, loads, and turbine control decisions are key factors impacting their economic effectiveness \cite{aderinto2018ocean}.}

\section*{Acknowledgments}
This work is supported in part by the National Science Foundation (Grants \#: ECCS-2114422 and ECCS-2114425).

\section*{Appendix A \\ Proof of Lemma 1} 
\begin{proof}
Scaling property of the Brownian Motion dictates that for any $s>0$, the process $\{ s^{-1/2} W(st) : t\geq 0\}$ is equivalent to a basis Brownian Motion process at time $t$. Evidently, it follows that $\int_0^t\sigma\cdot\left[\Psi\big(\phi(z;\tilde{\boldsymbol{\gamma}},\nu)\big)\right]^{\frac{1}{2}}dW(z)$ becomes equivalent to $\sigma \cdot W(\tau(t))$, where $\tau(t) = \int_0^t \left[\Psi\big(\phi(z;\tilde{\boldsymbol{\gamma}},\nu)\big)\right]dz$. Similarly, $\int_0^t \beta^0\cdot \Psi\big(\phi(z;\tilde{\boldsymbol{\gamma}},\nu)\big)dz = \int_0^{\tau(t)} \beta^0$. Given the equivalence of drift and Brownian error terms, $A(t)$ becomes equivalent to $A^0(\tau(t))$. Recall that, we define the RUL $\lambda$ as the first time period that the degradation signal $A(t)$ reaches a failure threshold $\Lambda$, i.e. $\lambda = min\left( t | A(t) \geq \Lambda \right)$. Given the above definition and equivalence of $A(t)$ and $A^0(\tau(t))$, the probability of survival until time $t$ can be written as follows: $P(\lambda>t) = P\left( sup_{0 \leq s\leq t} \left\{ A(t) \right\} \leq \Lambda \right) = P\left( sup_{0 \leq s\leq \tau(t)} \left\{ A^0(t) \right\} \leq \Lambda \right)$. The final expression is the first passage time of a Brownian Motion with constant drift $\beta^0$, which at any time $t$ follows an Inverse Gaussian distribution at the corresponding time-transformed period $\tau(t)$, namely, $\lambda \sim \mathcal{N}^{-1}\big(\tau(t)|a^0,b^0\big), t\geq0$, with $a^0=\frac{\Lambda-d_{t^c}}{\mu_{\beta^0}}$, and $b^0=\big(\frac{\Lambda-d_{t^c}}{\sigma}\big)^2$. \end{proof}

\section*{Appendix B: \\ The Dynamic Maintenance Cost Function}  \label{S:DMC}
The DMC function, first proposed in \cite{gebraeel2006sensory, yildirim2016sensor}, models the trade-off between conducting an early maintenance and wasting asset lifetime, versus delaying maintenance and risking asset failure. To do so, \eqref{eq:DMC_or} uses a renewal reward argument where it evaluates the cost within maintenance cycle in its numerator, and the expected length of the maintenance cycle in the denominator. In our context, the DMC, in its general form, can be expressed as in (\ref{eq:DMC_or}).
\begin{equation} \label{eq:DMC_or}
    c_{d,i} = \frac{C^{\text{PM}} \cdot P(\lambda_i>d) + C^{\text{CM}} \cdot P(\lambda_i\leq d)}{\int_0^d P(\lambda_i>z) dz + t_i^c},
\end{equation}
where $c_{d,i}$ represents the cost rate associated with conducting
maintenance $d$ time periods after the time of observation $t_i^c$. The numerator represents the expected maintenance cost, given the probability of failure $P(\lambda_i\leq d)$. The denominator represents the expected length of the cycle. Similarly, $c_i\in\mathbb{R}^+$ denotes the DMC of the $i$th WT. 
Note that the probability of failure, $P(\lambda_i\leq d)$ 
is a function of the YM decisions, as shown in (\ref{eq:yaw_degradation}), therefore, (\ref{eq:DMC_or}) is non-linear.

In what follows, we describe how we adapt the DMC in our framework. First, we re-write the integral in the denominator of (\ref{eq:DMC_or}) as: 
\begin{multline*}\int_0^d P(\lambda_i>z) dz = P(\lambda_i>1) + \int_1^d P(\lambda_i>z)dz \\ = P(\lambda_i>1) + \sum_{\tilde{d}=1}^d P(\lambda_i>\tilde{d}).
\end{multline*}
Given the assumption made for constraints (\ref{eq:zeta1})-(\ref{eq:zeta2L}), the probability of WT $i$ being operational at any day in our scenario-based framework can be written as $P(\lambda_i>1)=\sum_{s\in \mathcal{S} }\frac{\zeta_{i,s}}{N_{\mathcal{S}}}$ for the STH and as $P(\lambda_i>d)=\sum_{s\in \mathcal{S} }\frac{\zeta_{d,i,s}^L}{N_{\mathcal{S}}}=1-P(\lambda_i\leq d)$ for the LTH. 
By substitution, and by multiplying the numerator and the denominator by $N_{\mathcal{S}}$, we get the following expression for the DMC:
$$c_{d,i}^L = \frac{C^{\text{PM}} \cdot \sum_{s\in \mathcal{S}}\zeta_{d,i,s}^L + C^{\text{CM}} \cdot \left( N_{\mathcal{S}}- \sum_{s\in \mathcal{S}}\zeta_{d,i,s}^L \right )}{\sum_{s\in \mathcal{S} }\zeta_{i,s} + \sum_{\tilde{d}=1}^d \sum_{s\in \mathcal{S} }\zeta_{\tilde{d},i,s}^L + N_{\mathcal{S}} \cdot t_i^c},$$
or, equivalently:
\begin{multline*}\sum_{s\in \mathcal{S} }(\zeta_{i,s} \cdot c_{d,i}^L) + \sum_{\tilde{d}=1}^d \sum_{s\in \mathcal{S} } (\zeta_{\tilde{d},i,s}^L \cdot c_{d,i}^L) + c_{d,i}^L \cdot N_S \cdot t_i^c \\= C^{\text{PM}} \cdot \sum_{s\in \mathcal{S}}\zeta_{d,i,s}^L + C^{\text{CM}} \cdot \left(N_S- \sum_{s\in \mathcal{S}}\zeta_{d,i,s}^L \right ).
\end{multline*}

A non-linear sum of products between binary and continuous variable pairs is present at the left-hand side of the equation. We deal with these non-linear terms as follows. First, we introduce $\alpha_{0,d,i,s}^{c,L} \in \mathbb{R}^+$ and $\alpha_{\tilde{d},d,i,s}^{c,L} \in \mathbb{R}^+$, which are defined as $\alpha_{0,d,i,s}^{c,L} = \zeta_{i,s} \cdot c_{d,i}^L$ and $\alpha_{\tilde{d},d,i,s}^{c,L} = \zeta_{\tilde{d},i,s}^L \cdot c_{d,i}^L$, respectively. We can now write the linearized DMC function for each day $d$ in the LTH, as in constraint (\ref{eq:dcf_LTH}).
\begin{multline}\label{eq:dcf_LTH}
\sum_{s\in \mathcal{S}}\alpha_{0,d,i,s}^{c,L} + \sum_{\tilde{d}=1}^d \sum_{s\in \mathcal{S}} \alpha_{\tilde{d},d,i,s}^{c,L} + c_{d,i}^L \cdot N_S \cdot t_i^c \\= C^{\text{PM}} \cdot \sum_{s\in \mathcal{S}}\zeta_{d,i,s}^L + C^{\text{CM}} \cdot \left(N_S- \sum_{s\in \mathcal{S}}\zeta_{d,i,s}^L \right ) \quad \forall d \in \mathcal{D}, i \in \mathcal{I}
\end{multline}
It is followed by the set of constraints (\ref{dc_lin_4})-(\ref{dc_lin_9}), which are used to linearly control $\alpha_{0,d,i,s}^{c,L}$ and $\alpha_{\tilde{d},d,i,s}^{c,L}$. 
\begin{equation}\label{dc_lin_4}
\alpha_{0,d,i,s}^{c,L} \leq \mbox{M} \cdot \zeta_{i,s} \quad  \forall d \in \mathcal{D}, i \in \mathcal{I}, s\in \mathcal{S}
\end{equation}
\begin{equation}\label{dc_lin_5}
\alpha_{0,d,i,s}^{c,L} \leq c_{d,i}^L \quad  \forall  d \in \mathcal{D}, i \in \mathcal{I}, s\in \mathcal{S}
\end{equation}
\begin{equation}\label{dc_lin_6}
\alpha_{0,d,i,s}^{c,L} \geq c_{d,i}^L  - {\mbox{M}}\cdot(1-\zeta_{i,s}) \quad  \forall d \in \mathcal{D}, i \in \mathcal{I}, s\in \mathcal{S}
\end{equation}
\begin{multline}\label{dc_lin_7}
\alpha_{\tilde{d},d,i,s}^{c,L} \leq {\mbox{M}} \cdot \zeta_{\tilde{d},i,s}^L \quad  \forall \tilde{d} \in \{ 1, ..., d\}, d \in \mathcal{D}, i \in \mathcal{I}, \\s\in \mathcal{S}
\end{multline}
\begin{equation}\label{dc_lin_8}
\alpha_{\tilde{d},d,i,s}^{c,L} \leq c_{d,i}^L \quad  \forall \tilde{d} \in \{ 1, ..., d\},  d \in \mathcal{D}, i \in \mathcal{I}, s\in \mathcal{S}
\end{equation}
\begin{multline}\label{dc_lin_9}
\alpha_{\tilde{d},d,i,s}^{c,L} \geq c_{d,i}^L  - {\mbox{M}}\cdot(1-\zeta_{\tilde{d},i,s}^L) \quad  \forall \tilde{d} \in \{ 1, ..., d\}, d \in \mathcal{D}, \\i \in \mathcal{I}, s\in \mathcal{S}
\end{multline}

Following the same process, a linearized equivalent of the DMC function for the STH is described in constraint (\ref{dcf_STH}).
\begin{multline}\label{dcf_STH}
\sum_{s\in \mathcal{S} }\alpha_{i,s}^c + c_i \cdot N_S \cdot t_i^c = C^{\text{PM}} \cdot \sum_{s\in \mathcal{S} }\zeta_{i,s} + C^{\text{CM}} \cdot \left (N_S-\sum_{s\in \mathcal{S} }\zeta_{i,s} \right ) \\\quad \forall i \in \mathcal{I}
\end{multline}
We then use the set of linear constraints (\ref{dc_lin_1})-(\ref{dc_lin_3}) to control $\alpha_{i,s}^c\in\mathbb{R}^+$, which is defined as $\alpha_{i,s}^c=c_i\cdot \zeta_{i,s}$.
\begin{equation}\label{dc_lin_1}
\alpha_{i,s}^c \leq \mbox{M} \cdot \zeta_{i,s} \quad  \forall i \in \mathcal{I}, s\in \mathcal{S}
\end{equation}
\begin{equation}\label{dc_lin_2}
\alpha_{i,s}^c \leq c_i \quad  \forall i \in \mathcal{I}, s\in \mathcal{S}
\end{equation}
\begin{equation}\label{dc_lin_3}
\alpha_{i,s}^c \geq c_i - \mbox{M}\cdot(1-\zeta_{i,s}) \quad  \forall i \in \mathcal{I}, s\in \mathcal{S}
\end{equation}

{For additional demonstration of how we adapt the DMC in our framework, a video showing the change in the DMC function evaluations and the correspondent maintenance schedules from POSYDON over time is accessible at \cite{posydonvideo}.}

\section*{{Appendix C: \\ Pseudocode of POSYDON}}  \label{S:Alg}

\begin{algorithm}[]
\caption{{POSYDON Implementation Pseudocode}}\label{alg:alg1}
\begin{algorithmic}
\STATE \textit{Set} cost and operational parameters listed in Table \ref{tab:pars}
\STATE \textit{Input} historical weather, degradation, and operational data
\STATE \textit{Set} $N_{rolls}$ (here, $N_{rolls} = 238$ days; $1$ roll $= 1$ day)
\FOR{$r \in  \{1, ..., N_{rolls}\}$}
\STATE \textit{Train} probabilistic forecast models for $\nu$, $\eta$, $\kappa$, and $\lambda^0$ 
\STATE \textit{Sample}  $N_{\mathcal{S}}$ scenarios for $v$, $\eta$, $\kappa$, and $\lambda^0$
\STATE \textit{Evaluate} yaw-dependent power output, $f_{t,i,j,s}$ and $f^L_{d,i,j,s}$ 
\STATE \textit{Evaluate} relative RUL change, $F_{t,i,j,s}$ and $F^L_{d,i,j,s}$
\STATE \textit{Evaluate} expected mission times, $B_{t,i,s}$ and $B^L_{d,i,s}$
\STATE \textit{Solve} POSYDON (Optim. Gap$\leftarrow0.1$\%; Limit$\leftarrow30$ min) 
\STATE \textit{Return} the decisions for the $r$th roll: $\{\mathbf{m}_{r}, \pmb{\gamma}_{r}, \mathbf{m}^L_{r}, \pmb{\gamma}^L_{r}\}$
\STATE \textit{Shift} horizon by $1$ day, reveal new weather and operational data, and update the true RULs given executed decisions and newly revealed data 
\ENDFOR
\RETURN the final executed production and maintenance decisions, $\{\mathcal{M}^*, \Gamma^*\} := \{\mathbf{m}_r, \pmb{\gamma}_r\}_{r=1}^{N_{rolls}}$ 
\end{algorithmic}
\label{alg1}
\end{algorithm}

\bibliographystyle{IEEEtran}
\bibliography{references}

\begin{thebibliography}{10}
\providecommand{\url}[1]{#1}
\csname url@samestyle\endcsname
\providecommand{\newblock}{\relax}
\providecommand{\bibinfo}[2]{#2}
\providecommand{\BIBentrySTDinterwordspacing}{\spaceskip=0pt\relax}
\providecommand{\BIBentryALTinterwordstretchfactor}{4}
\providecommand{\BIBentryALTinterwordspacing}{\spaceskip=\fontdimen2\font plus
\BIBentryALTinterwordstretchfactor\fontdimen3\font minus
  \fontdimen4\font\relax}
\providecommand{\BIBforeignlanguage}[2]{{%
\expandafter\ifx\csname l@#1\endcsname\relax
\typeout{** WARNING: IEEEtran.bst: No hyphenation pattern has been}%
\typeout{** loaded for the language `#1'. Using the pattern for}%
\typeout{** the default language instead.}%
\else
\language=\csname l@#1\endcsname
\fi
#2}}
\providecommand{\BIBdecl}{\relax}
\BIBdecl

\bibitem{besnard2011stochastic}
F.~{Besnard}, M.~{Patriksson}, A.~{Strömberg}, A.~{Wojciechowski},
  K.~{Fischer}, and L.~{Bertling}, ``A stochastic model for opportunistic
  maintenance planning of offshore wind farms,'' in \emph{2011 IEEE Trondheim
  PowerTech}.\hskip 1em plus 0.5em minus 0.4em\relax IEEE, 2011, pp. 1--8.

\bibitem{yildirim2017integrated}
M.~Yildirim, N.~Gebraeel, and A.~Sun, ``Integrated predictive analytics and
  optimization for opportunistic maintenance and operations in wind farms,''
  \emph{IEEE Transactions on Power Systems}, vol.~32, no.~6, pp. 4319--4328,
  2017.

\bibitem{fallahi2022chance}
F.~Fallahi, I.~Bakir, M.~Yildirim, and Z.~Ye, ``A chance-constrained
  optimization framework for wind farms to manage fleet-level availability in
  condition based maintenance and operations,'' \emph{Renewable and Sustainable
  Energy Reviews}, vol. 168, p. 112789, 2022.

\bibitem{papadopoulos2022seizing}
P.~Papadopoulos, D.~W. Coit, and A.~A. Ezzat, ``Seizing opportunity:
  Maintenance optimization in offshore wind farms considering accessibility,
  production, and crew dispatch,'' \emph{IEEE Transactions on Sustainable
  Energy}, vol.~13, no.~1, pp. 111--121, 2022.

\bibitem{ding2012opportunistic}
F.~Ding and Z.~Tian, ``Opportunistic maintenance for wind farms considering
  multi-level imperfect maintenance thresholds,'' \emph{Renewable Energy},
  vol.~45, pp. 175--182, 2012.

\bibitem{byon2010optimal}
E.~Byon, L.~Ntaimo, and Y.~Ding, ``Optimal maintenance strategies for wind
  turbine systems under stochastic weather conditions,'' \emph{IEEE
  Transactions on Reliability}, vol.~59, no.~2, pp. 393--404, 2010.

\bibitem{byon2010season}
E.~Byon and Y.~Ding, ``Season-dependent condition-based maintenance for a wind
  turbine using a partially observed markov decision process,'' \emph{IEEE
  Transactions on Power Systems}, vol.~25, no.~4, pp. 1823--1834, 2010.

\bibitem{alvarez2022power}
M.~S. Alvarez-Alvarado, D.~L. Donaldson, A.~A. Recalde, H.~H. Noriega, Z.~A.
  Khan, W.~Velasquez, and C.~D. Rodriguez-Gallegos, ``Power system reliability
  and maintenance evolution: {A} critical review and future perspectives,''
  \emph{IEEE Access}, vol.~10, pp. 51\,922--51\,950, 2022.

\bibitem{yang2021review}
J.~Yang, L.~Fang, D.~Song, M.~Su, X.~Yang, L.~Huang, and Y.~H. Joo, ``Review of
  control strategy of large horizontal-axis wind turbines yaw system,''
  \emph{Wind Energy}, vol.~24, no.~2, pp. 97--115, 2021.

\bibitem{gebraad2013model}
P.~M. Gebraad, F.~C. van Dam, and J.-W. van Wingerden, ``A model-free
  distributed approach for wind plant control,'' in \emph{2013 American control
  conference}.\hskip 1em plus 0.5em minus 0.4em\relax IEEE, 2013, pp. 628--633.

\bibitem{johnson2012assessment}
K.~E. Johnson and G.~Fritsch, ``Assessment of extremum seeking control for wind
  farm energy production,'' \emph{Wind Engineering}, vol.~36, no.~6, pp.
  701--715, 2012.

\bibitem{fleming2014evaluating}
P.~A. Fleming, P.~M. Gebraad, S.~Lee, J.-W. van Wingerden, K.~Johnson,
  M.~Churchfield, J.~Michalakes, P.~Spalart, and P.~Moriarty, ``Evaluating
  techniques for redirecting turbine wakes using {SOWFA},'' \emph{Renewable
  Energy}, vol.~70, pp. 211--218, 2014.

\bibitem{kragh2015potential}
K.~A. Kragh and M.~H. Hansen, ``Potential of power gain with improved yaw
  alignment,'' \emph{Wind Energy}, vol.~18, no.~6, pp. 979--989, 2015.

\bibitem{song2018maximum}
D.~Song, J.~Yang, X.~Fan, Y.~Liu, A.~Liu, G.~Chen, and Y.~H. Joo, ``Maximum
  power extraction for wind turbines through a novel yaw control solution using
  predicted wind directions,'' \emph{Energy conversion and management}, vol.
  157, pp. 587--599, 2018.

\bibitem{howland2019wind}
M.~F. Howland, S.~K. Lele, and J.~O. Dabiri, ``Wind farm power optimization
  through wake steering,'' \emph{Proceedings of the National Academy of
  Sciences}, vol. 116, no.~29, pp. 14\,495--14\,500, 2019.

\bibitem{papadopoulos2022stochos}
P.~Papadopoulos, D.~W. Coit, and A.~A. Ezzat, ``{STOCHOS}: Stochastic
  opportunistic maintenance scheduling for offshore wind farms,'' \emph{IISE
  Transactions}, 2022, in Press.

\bibitem{yang2020operations}
L.~Yang, G.~Li, Z.~Zhang, X.~Ma, and Y.~Zhao, ``Operations \& maintenance
  optimization of wind turbines integrating wind and aging information,''
  \emph{IEEE Transactions on Sustainable Energy}, vol.~12, no.~1, pp. 211--221,
  2020.

\bibitem{yang2020operations2}
L.~Yang, R.~Peng, G.~Li, and C.-G. Lee, ``Operations management of wind farms
  integrating multiple impacts of wind conditions and resource constraints,''
  \emph{Energy Conversion and Management}, vol. 205, p. 112162, 2020.

\bibitem{erguido2017dynamic}
A.~Erguido, A.~C. M{\'a}rquez, E.~Castellano, and J.~G. Fern{\'a}ndez, ``A
  dynamic opportunistic maintenance model to maximize energy-based availability
  while reducing the life cycle cost of wind farms,'' \emph{Renewable Energy},
  vol. 114, pp. 843--856, 2017.

\bibitem{mazidi2017strategic}
P.~Mazidi, Y.~Tohidi, and M.~A. Sanz-Bobi, ``Strategic maintenance scheduling
  of an offshore wind farm in a deregulated power system,'' \emph{Energies},
  vol.~10, no.~3, p. 313, 2017.

\bibitem{kragh2014load}
K.~A. Kragh and M.~H. Hansen, ``Load alleviation of wind turbines by yaw
  misalignment,'' \emph{Wind Energy}, vol.~17, no.~7, pp. 971--982, 2014.

\bibitem{damiani2018assessment}
R.~Damiani, S.~Dana, J.~Annoni, P.~Fleming, J.~Roadman, J.~van Dam, and
  K.~Dykes, ``Assessment of wind turbine component loads under yaw-offset
  conditions,'' \emph{Wind Energy Science}, vol.~3, no.~1, pp. 173--189, 2018.

\bibitem{van2016yaw}
M.~T. van Dijk, J.-W. van Wingerden, T.~Ashuri, Y.~Li, and M.~A. Rotea,
  ``Yaw-misalignment and its impact on wind turbine loads and wind farm power
  output,'' in \emph{Journal of Physics: Conference Series}, vol. 753,
  no.~6.\hskip 1em plus 0.5em minus 0.4em\relax IOP Publishing, 2016, p.
  062013.

\bibitem{hellemo2018decision}
L.~Hellemo, P.~I. Barton, and A.~Tomasgard, ``Decision-dependent probabilities
  in stochastic programs with recourse,'' \emph{Computational Management
  Science}, vol.~15, pp. 369--395, 2018.

\bibitem{rezamand2020critical}
M.~Rezamand, M.~Kordestani, R.~Carriveau, D.~S.-K. Ting, M.~E. Orchard, and
  M.~Saif, ``Critical wind turbine components prognostics: A comprehensive
  review,'' \emph{IEEE Transactions on Instrumentation and Measurement},
  vol.~69, no.~12, pp. 9306--9328, 2020.

\bibitem{carroll2016failure}
J.~Carroll, A.~McDonald, and D.~McMillan, ``Failure rate, repair time and
  unscheduled {O\&M} cost analysis of offshore wind turbines,'' \emph{Wind
  Energy}, vol.~19, no.~6, pp. 1107--1119, 2016.

\bibitem{ennis2018wind}
B.~L. Ennis, J.~R. White, and J.~A. Paquette, ``Wind turbine blade load
  characterization under yaw offset at the {SWiFT} facility,'' in \emph{Journal
  of Physics: Conference Series}, vol. 1037, no.~5.\hskip 1em plus 0.5em minus
  0.4em\relax IOP Publishing, 2018, p. 052001.

\bibitem{bergami2014analysis}
L.~Bergami and M.~Gaunaa, ``Analysis of aeroelastic loads and their
  contributions to fatigue damage,'' in \emph{Journal of Physics: Conference
  Series}, vol. 555, no.~1.\hskip 1em plus 0.5em minus 0.4em\relax IOP
  Publishing, 2014, p. 012007.

\bibitem{hure2015optimal}
N.~Hure, R.~Turnar, M.~Va{\v{s}}ak, and G.~Ben{\v{c}}i{\'c}, ``Optimal wind
  turbine yaw control supported with very short-term wind predictions,'' in
  \emph{2015 IEEE International Conference on Industrial Technology
  (ICIT)}.\hskip 1em plus 0.5em minus 0.4em\relax IEEE, 2015, pp. 385--391.

\bibitem{zhang2018degradation}
Z.~Zhang, X.~Si, C.~Hu, and Y.~Lei, ``Degradation data analysis and remaining
  useful life estimation: A review on {W}iener-process-based methods,''
  \emph{European Journal of Operational Research}, vol. 271, no.~3, pp.
  775--796, 2018.

\bibitem{gebraeel2006sensory}
N.~{Gebraeel}, ``Sensory-updated residual life distributions for components
  with exponential degradation patterns,'' \emph{IEEE Transactions on
  Automation Science and Engineering}, vol.~3, no.~4, pp. 382--393, 2006.

\bibitem{zhou2016novel}
Y.~Zhou, M.~Huang, Y.~Chen, and Y.~Tao, ``A novel health indicator for on-line
  lithium-ion batteries remaining useful life prediction,'' \emph{Journal of
  Power Sources}, vol. 321, pp. 1--10, 2016.

\bibitem{zhai2017rul}
Q.~Zhai and Z.-S. Ye, ``{RUL} prediction of deteriorating products using an
  adaptive {W}iener process model,'' \emph{IEEE Transactions on Industrial
  Informatics}, vol.~13, no.~6, pp. 2911--2921, 2017.

\bibitem{koltsidopoulos2020data}
A.~Koltsidopoulos~Papatzimos, ``Data-driven operations \& maintenance for
  offshore wind farms: Tools and methodologies,'' Ph.D. dissertation,
  University of Exeter, 2020.

\bibitem{scholbrock2014field}
A.~Scholbrock, P.~Fleming, A.~Wright, C.~Slinger, J.~Medley, and M.~Harris,
  ``Field test results from lidar measured yaw control for improved yaw
  alignment with the {NREL} controls advanced research turbine,'' National
  Renewable Energy Lab.(NREL), Golden, CO (United States), Tech. Rep., 2014.

\bibitem{pranjaal2022yaw}
P.~Nasery and A.~A. Ezzat, ``Yaw-adjusted wind power curve modeling: A local
  regression approach,'' \emph{Renewable Energy}, 2022, in Press.

\bibitem{golparvar2021surrogate}
B.~Golparvar, P.~Papadopoulos, A.~Ezzat, and R.~Wang, ``A surrogate-model-based
  approach for estimating the first and second-order moments of offshore wind
  power,'' \emph{Applied Energy}, vol. 299, p. 117286, 2021.

\bibitem{ding2019data}
Y.~Ding, \emph{Data Science for Wind Energy}.\hskip 1em plus 0.5em minus
  0.4em\relax New York: Chapman and Hall/CRC, 2019.

\bibitem{lease2021}
BOEM, ``Lease and grant information,'' 2021, {B}ureau of Ocean Energy
  Management. Available at:
  \url{https://www.boem.gov/renewable-energy/lease-and-grant-information}.

\bibitem{nyserda2019}
NYSERDA, ``{NYSERDA} floating {L}i{DAR} buoy data,'' 2021, {N}ew {Y}ork {S}tate
  Energy Research \& Development Agency. Available at:
  \url{https://oswbuoysny.resourcepanorama.dnvgl.com}.

\bibitem{optis2020validation}
M.~Optis, A.~Kumler, G.~N. Scott, M.~C. Debnath, and P.~J. Moriarty,
  ``Validation of {RU-WRF}, the custom atmospheric mesoscale model of the
  {R}utgers {C}enter for {O}cean {O}bserving {L}eadership,'' NREL, Golden, CO,
  USA, Tech. Rep., 2020.

\bibitem{wavewatch}
NOAA, ``{WAVEWATCH III} model description,'' 2022, {N}ational Oceanic and
  Atmospheric Administration. Available at:
  \url{https://polar.ncep.noaa.gov/waves/wavewatch/}.

\bibitem{PJM}
PJM, ``{PJM} data directory,'' 2021, available at:
  \url{https://dataminer2.pjm.com}.

\bibitem{lago2021forecasting}
J.~Lago, G.~Marcjasz, B.~De~Schutter, and R.~Weron, ``Forecasting day-ahead
  electricity prices: A review of state-of-the-art algorithms, best practices
  and an open-access benchmark,'' \emph{Applied Energy}, vol. 293, p. 116983,
  2021.

\bibitem{ezzat2018spatio}
A.~Ezzat, M.~Jun, and Y.~Ding, ``Spatio-temporal asymmetry of local wind fields
  and its impact on short-term wind forecasting,'' \emph{IEEE Transactions on
  Sustainable Energy}, vol.~9, no.~3, pp. 1437--1447, 2018.

\bibitem{ye2023airu}
F.~Ye, J.~Brodie, T.~Miles, and A.~A. Ezzat, ``{AIRU-WRF}: A physics-guided
  spatio-temporal wind forecasting model and its application to the {US}
  {N}orth {A}tlantic offshore wind energy areas,'' \emph{arXiv preprint
  arXiv:2303.02246}, 2023.

\bibitem{gebraeel2005residual}
N.~Z. Gebraeel, M.~A. Lawley, R.~Li, and J.~K. Ryan, ``Residual-life
  distributions from component degradation signals: A {B}ayesian approach,''
  \emph{IIE Transactions}, vol.~37, no.~6, pp. 543--557, 2005.

\bibitem{ezzat2019spatio}
A.~Ezzat, M.~Jun, and Y.~Ding, ``Spatio-temporal short-term wind forecast: A
  calibrated regime-switching method,'' \emph{The Annals of Applied
  Statistics}, vol.~13, no.~3, pp. 1484--1510, 2019.

\bibitem{ezzat2020turbine}
A.~Ezzat, ``Turbine-specific short-term wind speed forecasting considering
  within-farm wind field dependencies and fluctuations,'' \emph{Applied
  Energy}, vol. 269, p. 115034, 2020.

\bibitem{aderinto2018ocean}
T.~Aderinto and H.~Li, ``Ocean wave energy converters: Status and challenges,''
  \emph{Energies}, vol.~11, no.~5, p. 1250, 2018.

\bibitem{yildirim2016sensor}
M.~Yildirim, X.~A. Sun, and N.~Z. Gebraeel, ``Sensor-driven condition-based
  generator maintenance scheduling—{Part} {I}: Maintenance problem,''
  \emph{IEEE Transactions on Power Systems}, vol.~31, no.~6, pp. 4253--4262,
  2016.

\bibitem{posydonvideo}
\BIBentryALTinterwordspacing
P.~Papadopoulos and A.~Ezzat, ``Research {M}ultimedia: {POSYDON}: Balancing the
  short- and long-term needs of offshore wind energy operation,''
  \textit{{L}ast Accessed: June, 2023}. [Online]. Available:
  \url{https://sites.rutgers.edu/azizezzat/research-videos-multimedia/}
\BIBentrySTDinterwordspacing

\end{thebibliography}


\begin{IEEEbiography}[{\includegraphics[width=1in,height=1.25in,clip,keepaspectratio]{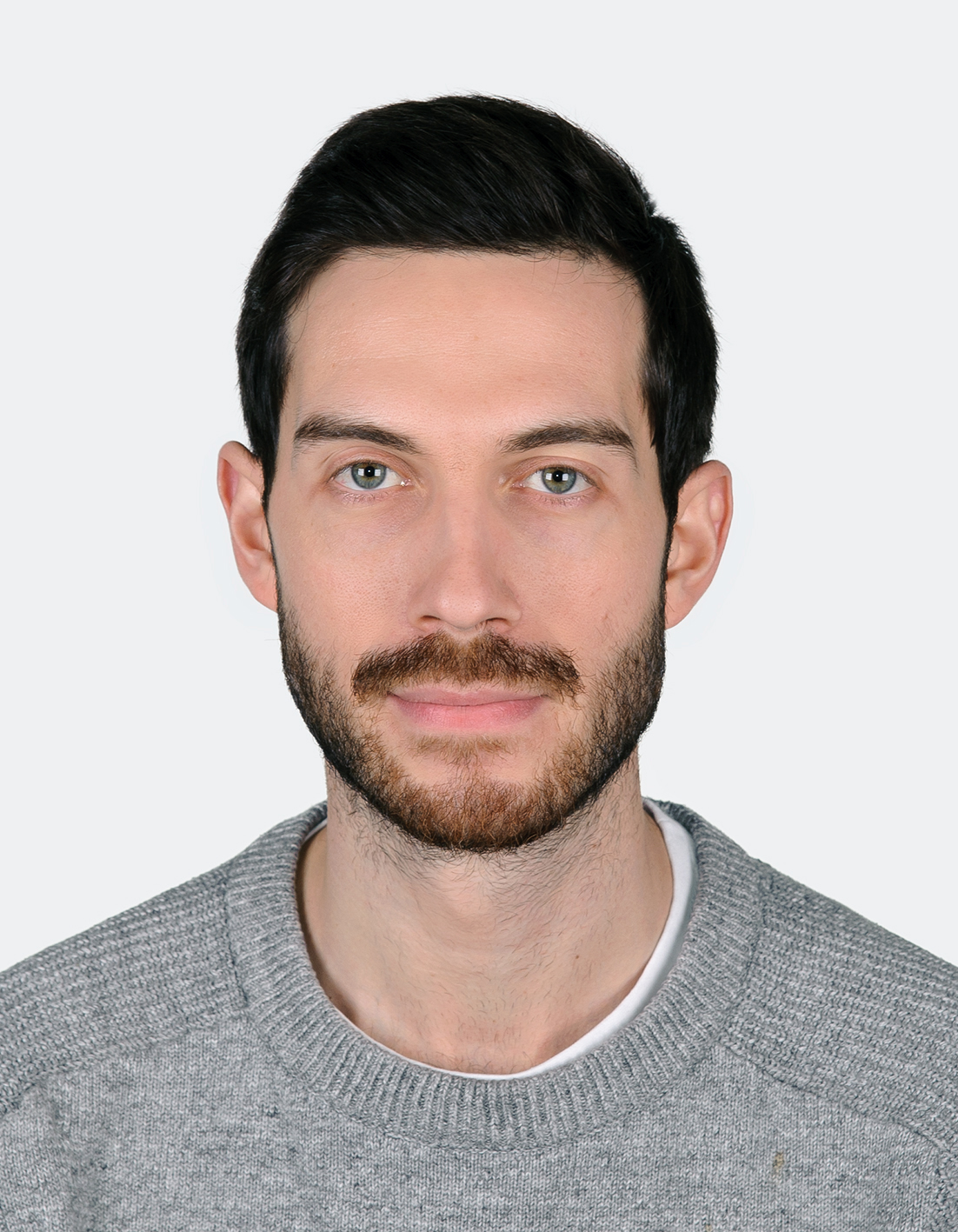}}]{Petros Papadopoulos}
received his Ph.D. in Industrial and Systems Engineering from Rutgers University in New Jersey, USA in 2022, and an M.Eng. degree in Chemical Engineering from the University of Patras in Greece in 2018. His research interests encompass data analytics and mathematical optimization for energy systems and electricity market services. He is currently working as a Research and Development Engineer at ThermoVault in Leuven, Belgium, focusing on demand response services for decentralized storage systems.
\end{IEEEbiography}

\begin{IEEEbiography}[{\includegraphics[width=1in,height=1.25in,clip,keepaspectratio]{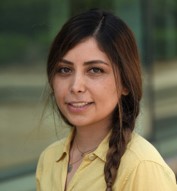}}]{Farnaz Fallahi} is currently pursuing a Ph.D. degree in Industrial and Systems Engineering at Wayne State University, Detroit, MI, USA. She obtained her B.S. degree in Pure Mathematics from Amirkabir University of Technology, Tehran, Iran, in 2011, and her M.S. degree in Industrial and Systems Engineering from Sharif University of Technology, Tehran, Iran, in 2015. Her research interests include sensor-driven prognostics, stochastic programming, and data-driven decision-making, with a specific focus on energy systems and complex networks.
\end{IEEEbiography}

\begin{IEEEbiography}[{\includegraphics[width=1in,height=1.25in,clip,keepaspectratio]{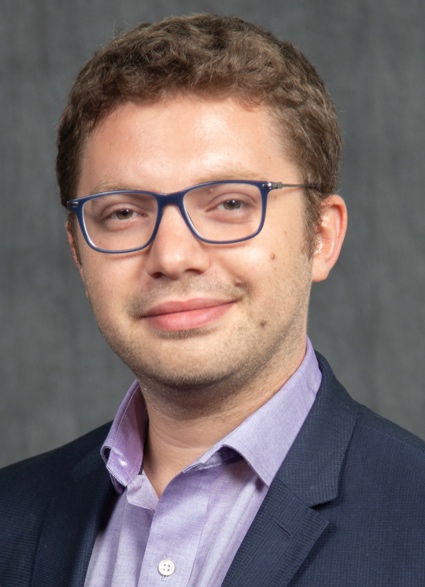}}]{Murat Yildirim} is an assistant professor in the Department of Industrial and Systems Engineering, and the Director of Cyber Physical Systems Laboratory at Wayne State University. He obtained a PhD degree in Industrial Engineering, MSc degree in Operations Research, and BSc degrees in Electrical and Industrial Engineering from Georgia Institute of Technology. Dr. Yildirim's research interest lies on the modeling and the computational challenges arising from the integration of real-time inferences generated by advanced data analytics and simulation into large-scale mathematical programming models used for optimizing and controlling networked systems. His research has been supported through multiple projects funded by NSF, DoE, MTRAC and Ford.
\end{IEEEbiography}

\begin{IEEEbiography}[{\includegraphics[width=1in,height=1.25in,clip,keepaspectratio]{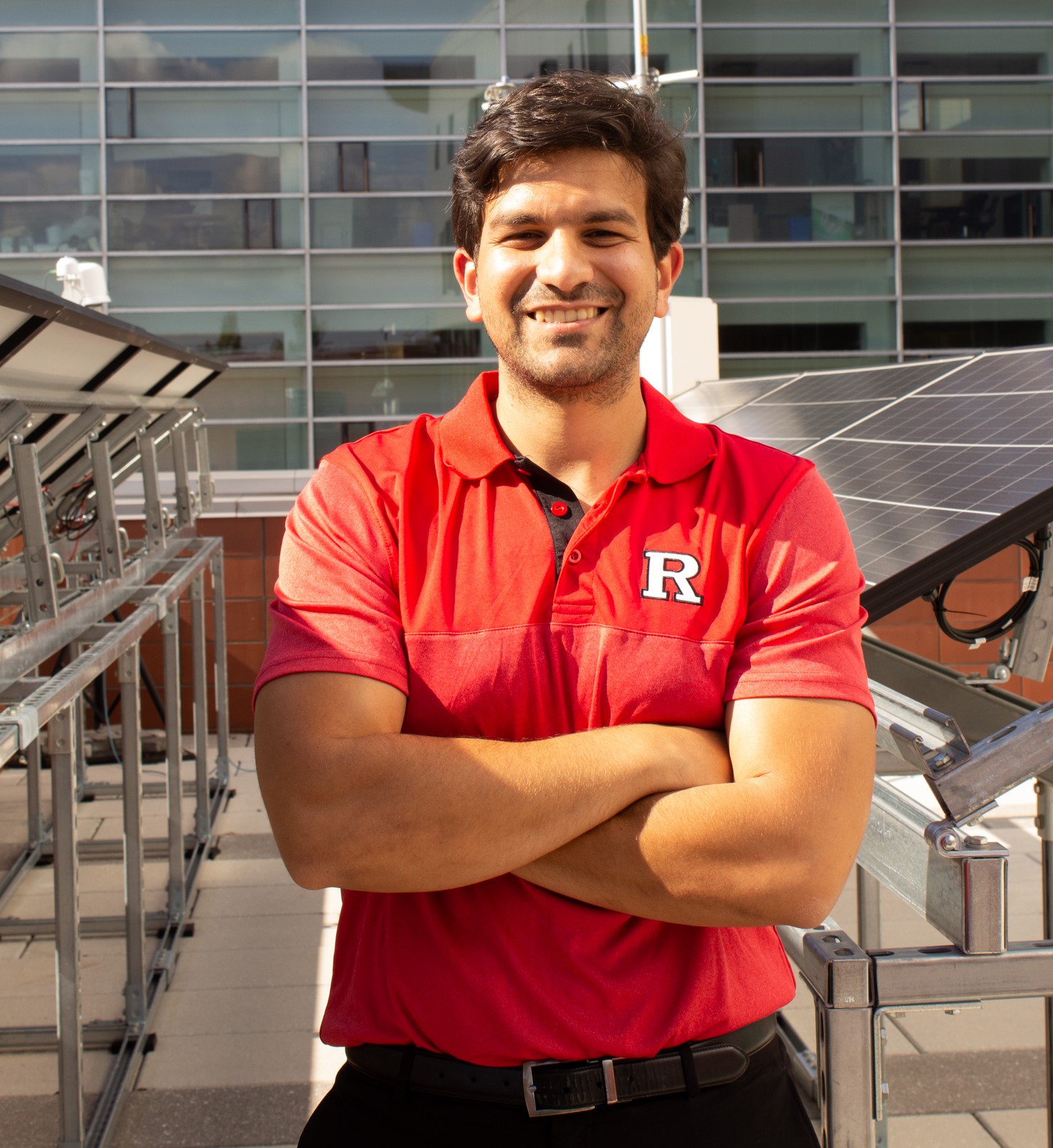}}]{Ahmed Aziz Ezzat} is an Assistant Professor of Industrial \& Systems Engineering at Rutgers University, NJ, where he leads the Renewables \& Industrial Analytics (RIA) research group. Before joining Rutgers, Dr. Aziz Ezzat received his Ph.D. from Texas A\&M University in 2019, and his B.Sc. degree from Alexandria, Egypt, in 2013, both in Industrial Engineering. His research interests are in the areas of data and decision sciences, forecasting analytics, quality/reliability engineering, with focus on renewable energy systems and industrial informatics. 
His research has been supported by NSF, NOWRDC, NJEDA, as well as industry. He is currently serving as the president of the IISE Energy Systems division and the secretary of the Forecasting for Social Good (F4SG) cluster at IIF. He is a member of INFORMS, IISE, and IEEE-PES. 
\end{IEEEbiography}

\vfill

\end{document}